\documentclass[usenatbib,useAMS]{mn2e}

\usepackage{epsf}
\usepackage[dvips]{epsfig}
\usepackage{times}

\begin{document}

\label{firstpage}

\title[Enhanced $M/L$-ratios in UCDs]{Enhanced Mass-to-Light Ratios in
  UCDs through Tidal Interaction with the Centre of the Host Galaxy}

\author[M. Fellhauer and P. Kroupa] {M. Fellhauer$^{1,2,3}$
  \thanks{madf@ast.cam.ac.uk} and P. Kroupa$^{1,2}$ \thanks
  {pavel@astro.uni-bonn.de} \\
  $^{1}$ Argelander Institute for Astronomy, University Bonn, Auf dem
  H\"{u}gel 71, 53121 Bonn\\
  $^{2}$ The Rhine Stellar-Dynamical Network \\
  $^{3}$ Institute of Astronomy, University of Cambridge, Madingley Road, 
  Cambridge CB3 0HA} 

\pagerange{\pageref{firstpage}--\pageref{lastpage}} \pubyear{2005}

\maketitle

\begin{abstract}
  A recent study of ultra-compact dwarf galaxies (UCDs) in the Virgo
  cluster revealed that some of them show faint envelopes and have
  measured mass-to-light ratios of $5$ and larger, which can not be
  explained by simple population synthesis models.  It is believed
  that this proves that some of the UCDs must possess a dark matter
  halo and may therefore be stripped nuclei of dwarf ellipticals
  rather than merged star cluster complexes.

  Using an efficient N-body method we investigate if a close passage
  of a UCD through the central region of the host galaxy is able to
  enhance the measured mass-to-light ratio by tidal forces leaving the
  satellite slightly out of virial equilibrium and thereby leading to
  an overestimation of its virial mass.

  We find this to be possible and discuss the general problem of
  measuring dynamical masses for objects that are probably interacting
  with their hosts.
\end{abstract}

\begin{keywords}
  galaxies: dwarfs -- galaxies: interactions -- galaxies: kinematics
  and dynamics -- methods: N-body simulations
\end{keywords}

\section{Introduction}
\label{sec:intro}

\citet{has05} investigated ultra-compact dwarf galaxies (UCDs) around
M87, the central galaxy of the Virgo cluster.  By measuring the
surface brightness profiles and assessing the projected velocity
dispersion of the objects they concluded that some UCDs of their
sample have mass-to-light ratios of the order $5$--$9$.  Furthermore,
they find that some of the UCDs show faint envelopes.  This supports
the notion that these UCDs may be stripped nuclei of dwarf
ellipticals. 

UCDs were first discovered by \citet*{hil98,hil99} during a study of 
globular clusters and dwarf galaxies around the central galaxy in the
Fornax cluster.  These objects are compact with effective radii of
about $15$--$25$~pc and a hundred to several hundred pc in extension,
and they are massive with masses of a few $10^{6}$ up to a few
$10^{7}$~M$_{\odot}$.

There are several theories about the origin of UCDs: (I) They could be
the most luminous end of the distribution function of very massive
globular clusters \citep{hil99,mie02,dir03}.  (II) They could be the
remnants of stripped dwarf ellipticals \citep*{bek01,bek03,dep05}.  In
this 'threshing' scenario a nucleated dwarf elliptical looses its
envelope and most of its dark matter content due to tidal interaction
with the host galaxy, such that only the 'naked' nucleus remains.
(III) They could be amalgamated young massive star clusters formed in
a star cluster complex during the star-burst caused by the tidal
perturbation and possible disruption of a gas-rich galaxy
\citep{fel02,fel05}.  This scenario is well-established theoretically
and was first proposed by \citet{kro98}.  It must have been
profusively active during the early hierarchical structure formation
epoch when gas-rich substructures merged to the present-day major
galaxies.

\begin{table*}
  \centering
  \begin{minipage}{10.5cm}
    \caption{Table of our initial model parameters.  The columns
      denote the initial mass of our model ($M_{\rm pl} = M_{\rm
        ini}$), the scale length (Plummer radius, $R_{\rm pl}$;
      analytical \& measured as described in Sect.~\ref{sec:res}), the
      characteristic crossing time ($T_{\rm cr}$), the central
      projected (line-of-sight) velocity dispersion ($\sigma_{0,p}$;
      analytical \& measured as described in Sect.~\ref{sec:res}), and
      the scaling factor ($A$) to compute the virial mass (see
      Eq.~\ref{eq:virplum}).}
    \label{tab:para}
    \begin{tabular}[t!]{rrrrrrr} \hline Mass [M$_{\odot}$] & $R_{\rm
        pl}$ [pc] & measured & $T_{\rm cr}$ [Myr] & $\sigma_{0,p}$
      [km\,s$^{-1}$] & measured & $A$ \\ \hline \hline $10^{7}$ & $25$
      & $23.7$ & $3.69$ & $15.92$ & $16.20$ &
      $1608$ \\
      $10^{7}$ & $50$ & $46.6$ & $10.43$ & $11.25$ & $11.71$ &
      $1565$ \\
      $10^{7}$ & $100$ & $94.6$ & $29.50$ & $7.96$ & $8.43$ &
      $1487$ \\
      $10^{7}$ & $250$ & $221.0$ &$116.61$ & $5.03$ & $6.30$ & $1140$
      \\ \hline $10^{8}$ & $25$ & $25.2$ & $1.17$ & $50.33$ & $50.2$ &
      $1573$ \\
      $10^{8}$ & $50$ & $47.4$ & $3.30$ & $35.59$ & $36.8$ &
      $1558$ \\
      $10^{8}$ & $100$ & $92.8$ & $9.33$ & $25.16$ & $26.9$ &
      $1489$ \\
      $10^{8}$ & $250$ & $213.0$ & $36.88$ & $15.92$ & $19.91$ &
      $1184$ \\ \hline
    \end{tabular}
  \end{minipage}
\end{table*}

It is still under debate whether the UCDs, which fill the gap between
globular clusters and dwarf galaxies, follow the fundamental plane
(effective radius / velocity dispersion -- total luminosity) relation
for globular clusters or for dwarf galaxies \citep*{has05,evs05}.
While, in the $M_{V}$--$\sigma_{0}$-plane \citep[see e.g.][their
Fig.~7]{has05}, they lie closer to the globular clusters, their
relation seems to rather follow the one of dwarf galaxies.
Furthermore they occupy exactly the space between dwarf ellipticals
and their nuclei.  This seems to point to formation theory (II).  But
the surface brightness profiles of the bright UCDs in Fornax are much
more extended (i.e.\ larger effective radii) compared with nuclei of
dEs \citep{dep05}.

On the other hand, \citet{mar04} found an intermediate age object (W3,
age $300$--$500$~Myr) in the merger remnant galaxy NGC~7252.  The mass
($M = 8 \cdot 10^{7}$~M$_{\odot}$), size ($r_{\rm eff} = 17.5$~pc, and
velocity dispersion of $45$~km\,s$^{-1}$) strongly suggest this object
to be a UCD rather than a globular cluster.  The age of this object,
which corresponds to the time elapsed since the major interaction,
unambiguously shows that this object can not be a stripped nucleus of
a dwarf elliptical.  A dwarf elliptical can not be stripped in a time
interval of only $500$~Myr.  \citet{fel05} showed that W3 could be the
merger object of a massive star cluster complex which was formed
during the interaction.  The subsequent merging of star clusters forms
an object with properties similar to those of W3.  The evolution of
the simulated super cluster shows that it transforms into a UCD such
as those found in Fornax \citep{hil99,phi01}, Abell~1689
\citep{mie04}, and Virgo \citep{has05,evs05}.  Moreover, \citet{fel05}
showed that, due to its high mass, the object was able to retain an
envelope of bound stars which initially were expelled from the
individual clusters during the merger process.  Thus, an envelope
around a UCD is not a proof of its cosmological origin as a dwarf
elliptical.  Still the puzzle of high mass-to-light ratios of some of
the UCDs in Virgo remains.  These mass-to-light ratios (5--9) are
found for the UCDs lying closest to the host galaxy.  This suggests
that deviations from virial equilibrium may play a role.

The reasoning is that dwarf galaxies are on radial rather than
circular orbits, if the UCDs formed as star-cluster complexes during
the merging of major gas-rich substructures.  This allows the
satellites to pass close to the galactic centre.  Tidal forces in the
central region of the host galaxy are rather strong.  Hence, the dwarf
object looses stars and subsequently departs from virial equilibrium.
Some of the lost stars (stars which are no longer bound to the object)
do not immediately leave the object or its vicinity but disperse
slowly along the orbit.  Thus, a line-of-sight velocity dispersion
measurement may be contaminated by these unbound stars which inflate
the velocity dispersion.  Furthermore, the gravitational shock of the
central passage leads to an expansion of the dwarf galaxy which is
later reversed again.  But still this expansion through tidal heating
may result in a measurable increase of the core radius once the dwarf
object is again outside the host galaxy.  Both of these effects may
lead to an overestimation of the measured dynamical mass resulting in
a higher mass-to-light ratio.  These effects are not new to the
astronomical community and are studied intensively by various authors
(e.g. \citep{kro97} for dwarf spheroidals, \citep{may01,may02} for
dwarf discs and dwarf spheroidals or \citep{bek03} for dwarf
ellipticals).  With this paper we want to extend these studies onto
massive (compared to globular clusters) and compact (compared to other
dwarf galaxies) objects like UCDs. 

It is definitely clear that these effects must be very strong if an
object comes close to the galactic centre.  But on the other hand
these very close passages may be highly unlikely.  By means of
numerical simulations we intend to find out which sets of orbits allow
an enhanced mass-to-light ratio, i.e.\ how close the UCD has to
approach to the centre of its host galaxy, for which we choose the
parameters of M87 to account for the UCDs with high mass-to-light
ratios found around the central galaxy of Virgo.

\section{Setup}
\label{sec:setup}

We model the parent galaxy as an analytical potential because on one
single passage dynamical friction for a dwarf galaxy of mass $M \leq
10^{8}$~M$_{\odot}$ affects the orbit to at most $2$--$3$~per cent,
estimated using Chandrasekhar's formula \citep{cha43} as described in
\citet{por02}.  We also look for only one central passage because
unbound stars disperse along the orbit and are lost at subsequent
passages.  Some of them might still be around the object after the
second or third central passage but definitely not for dozens of
orbits.

As a model for the analytical potential we choose the parameters for
M87, the central galaxy in the Virgo cluster, consisting of a
NFW-profile for the dark halo, a Hernquist profile (H) for the visible
matter (stars) and a central super-massive black hole (BH)
\citep*{mcl99,ves03,dim03}.  The density profile of the host galaxy is
\begin{eqnarray}
  \label{eq:pot}
  \rho_{\rm tot}(r) & = & \rho_{\rm NFW}(r) + \rho_{\rm H}(r)
  \nonumber \\
  & = & \frac{\rho_{0,{\rm NFW}} \ r_{\rm s,NFW}} {r \left( 1 +
      \frac{r}{r_{\rm s,NFW}} \right)^{2}} + \frac{M_{\rm H} \ r_{\rm
      s,H}} {2 \pi \ r \left( r_{\rm s,H} + r \right)^{3}},
\end{eqnarray}
with the following parameters,
\begin{eqnarray}
  \label{eq:param}
  \rho_{0,{\rm NFW}} & = & 3.17 \cdot 10^{-4} \ {\rm M}_{\odot}/{\rm
    pc}^{3}, \\
  r_{\rm s,NFW} & = & 560 \ {\rm kpc}, \\
  M_{\rm H} & = & 8.1 \cdot 10^{11} \ {\rm M}_{\odot}, \\
  r_{\rm s,H} & = & 5.1 \ {\rm kpc}, \\
  M_{\rm BH} & = & 3 \cdot 10^{9} \ {\rm M}_{\odot}.
\end{eqnarray}
The above parameters denote from top to bottom the characteristic
density and the scale-length of the NFW-profile, the total mass and
the scale-length of the Hernquist-profile and finally the mass of the
central super-massive black hole.

We model the UCD as a Plummer-sphere \citep{plu11} in the numerical
realisation described by \citet*{aar74},
\begin{eqnarray}
  \label{eq:plummer}
  \rho_{\rm pl}(r) & = & \frac{3M_{\rm pl}}{4\pi R_{\rm pl}^{3}}
  \left( 1 + \frac{r^{2}}{R_{\rm pl}^{2}}\right)^{-5/2},
\end{eqnarray}
with varying scale-lengths, $R_{\rm pl}$, and initial masses, $M_{\rm
  pl}$, because \citet{dep05} found a Plummer-profile to fit four out
of five UCDs in the Fornax cluster quite well.  Furthermore, the
Plummer model is analytically simple and the Plummer radius is not
only the scale-length of the model but also its half-light radius (the
projected radius from within which half of the light of the object is
emitted).

The masses of UCDs range between several million M$_{\odot}$ up to
several tens of millions.  So we choose for our models $10^{7}$ and
$10^{8}$~M$_{\odot}$ as initial masses ($M_{\rm pl} = M_{\rm ini}$).
We vary the initial scale-length of our models to be $R_{\rm pl} =
25$, $50$, $100$ and $250$~pc to determine the influence of the
concentration of the objects.  The cut-off radius ($R_{\rm lim}$; the
radius where we truncate the Plummer distribution of our object) of
all our models was kept constant at $500$~pc which is larger than the
initial tidal radius.  A detailed list of our model parameters can be
found in Table~\ref{tab:para}.  The Plummer spheres are modelled using
$10^{6}$ particles.  The object is set-up and integrated in isolation
until equilibrium, as measured by the constancy of the $90$~\%
Lagrangian radius, is reached \citep{kro97}.

The UCD model is then placed at a distance of $10$~kpc to the centre of
the host galaxy with no radial velocity.  This means that the effect
of the central passage is the most harmful possible because it is the
slowest possible.  The faster the passage would be (if the satellite
was to start further out) the less tidal influence the central passage
would have.  No UCD is found closer than $10$~kpc from the centre of
its host and therefore it acts as a minimum apogalacticon, which has
the strongest tidal effect possible.  To vary the minimum distance
(perigalacticon) we give our models different tangential velocities
which are listed in Tab.~\ref{tab:vel}.  We assume the problem to be
spherically symmetric, so we are able to place the trajectory of our
model in the $x$-$y$-plane of our simulation area without restricting
the problem.

\begin{table}
  \centering
  \caption{List of minimum distances and the corresponding tangential
    velocities at the start of the simulation.} 
  \label{tab:vel}
  \begin{tabular}[t!]{rr} \hline
    $D_{\rm min}$ [pc] & $v_{\rm tan}$ [km\,s$^{-1}$] \\ \hline \hline
    0 & 0.0 \\  
    50 & 6.0 \\
    100 & 10.9 \\
    150 & 15.9 \\
    250 & 25.4 \\
    500 & 48.1 \\
    1000 & 89.3 \\
    1500 & 126.0 \\
    2000 & 158.9 \\ \hline
  \end{tabular}
\end{table}

We use the particle-mesh code {\sc Superbox} to carry out the
simulations.  {\sc Superbox} has high-resolution sub-grids which stay
focused on the core of the dwarf object while it is moving through
the host galaxy.  The resolution of the innermost grid containing the
core is $3$~pc.  For a detailed description of the code see
\citet{fel00}. 

\section{Results}
\label{sec:res}

We carried out a parameter survey of 76 simulations covering different
dwarf galaxy objects and different orbits.  The parameter range is
shown in Tables~\ref{tab:para} and~\ref{tab:vel}.  For each parameter
set one simulation over two central passages is carried out.  For the
determination of our results we look at the satellite when it reaches
apogalacticon again after the first central passage.

\subsection{The analytical method}
\label{sec:theo}

The surface density profile of the satellite is fitted by a Plummer
profile of the form
\begin{eqnarray}
  \label{eq:plum-surf}
  \Sigma(r) & = & \Sigma_{0} \left( 1 + \frac{r^{2}} {R_{\rm
        pl}^{2}}\right)^{-2},
\end{eqnarray}
by applying a non-linear least-squares Marquardt-Levenberg algorithm.
From this procedure we take the fitted Plummer radius for our
determination of the virial mass.  It can be shown that the Plummer
radius exactly coincides with the half-light (projected half-mass)
radius of a Plummer sphere.

Furthermore we determine the line-of-sight velocity dispersion
profile.  In cases where it is possible, i.e.\ the profile is not too
contaminated by unbound stars, we again fit the Plummer profile for
the line-of-sight velocity dispersion,
\begin{eqnarray}
  \label{eq:plum-sig}
  \sigma_{\rm p}(r) & = & \sigma_{\rm 0,p} \left( 1 + \frac{r^{2}}
    {R_{\rm pl}^{2}}\right)^{-1/4}.
\end{eqnarray}

The projected velocity dispersion, as well as the surface density, is
measured along all three Cartesian coordinates in logarithmically
spaced, concentric rings centred on the object.  For the measurement
all stars within a certain distance $R_{\rm max}$ in front and behind
the centre of the object are taken into account.  The procedure is
done along all three Cartesian axes and we take the arithmetic mean
value (i.e.\ a mean profile), because we do not know which orientation
our objects have with respect to the observer.  Therefore effects may be
very strong measuring along the trajectory of the object but almost
not visible in the perpendicular direction.  Any random direction is
likely to yield an intermediate result.

For the surface density we always fit a Plummer profile even if the
object is completely destroyed and does not follow a Plummer
distribution at all.  But in most of our models the remaining object
is still fairly well represented by a Plummer profile even if it is on
the way to complete destruction.  In cases where we are not able to
fit the velocity dispersion profile with a Plummer profile (see e.g.\
first panel of Fig.~\ref{fig:rmax}) we take an average of all measured
line-of-sight velocity dispersion values within the innermost $10$~pc
in projected radius to determine the central line-of-sight velocity
dispersion.

\begin{figure}
    \centering 
    \epsfxsize=7cm 
    \epsfysize=7cm
    \epsffile{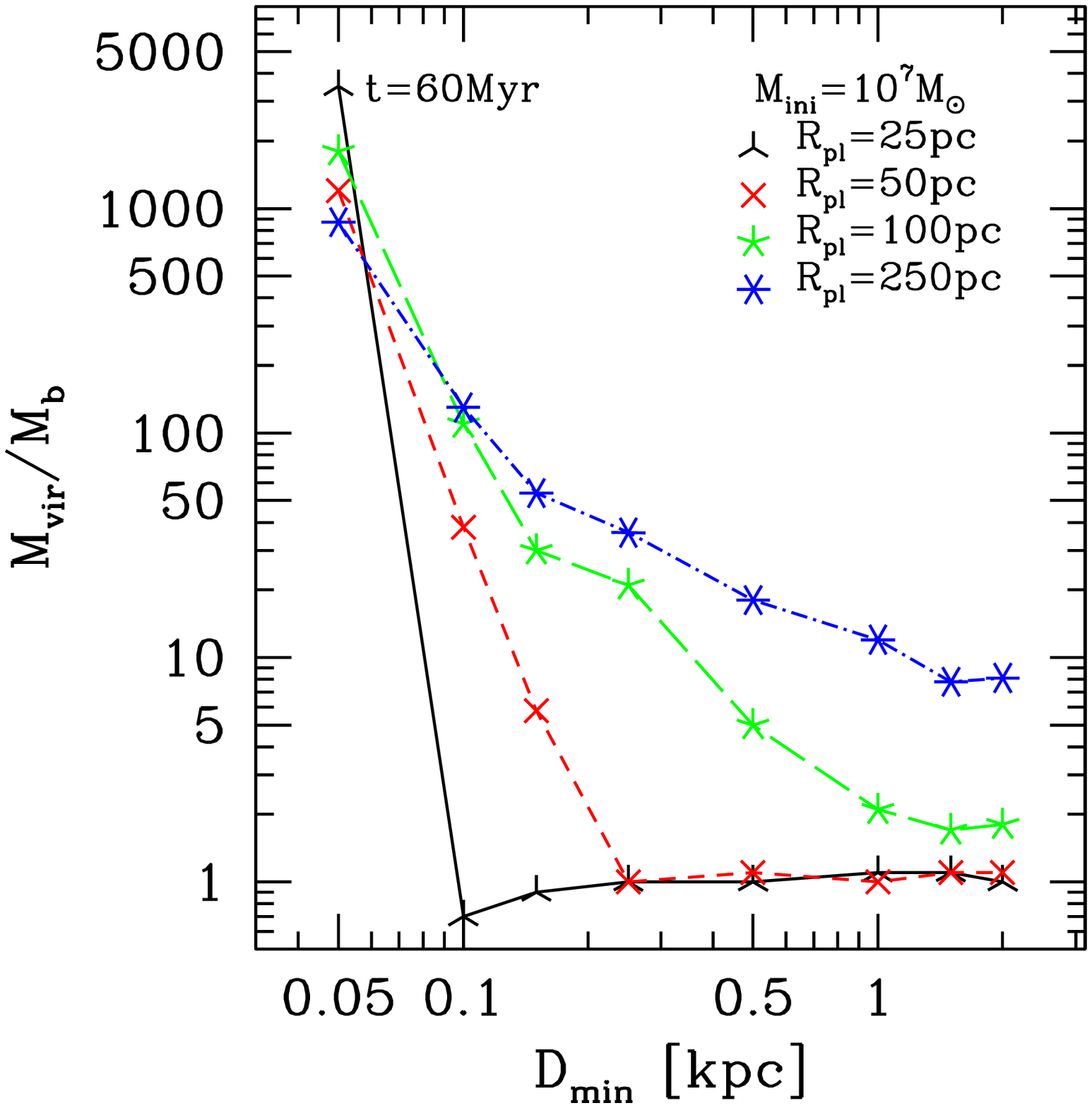}
    \epsfxsize=7cm 
    \epsfysize=7cm 
    \epsffile{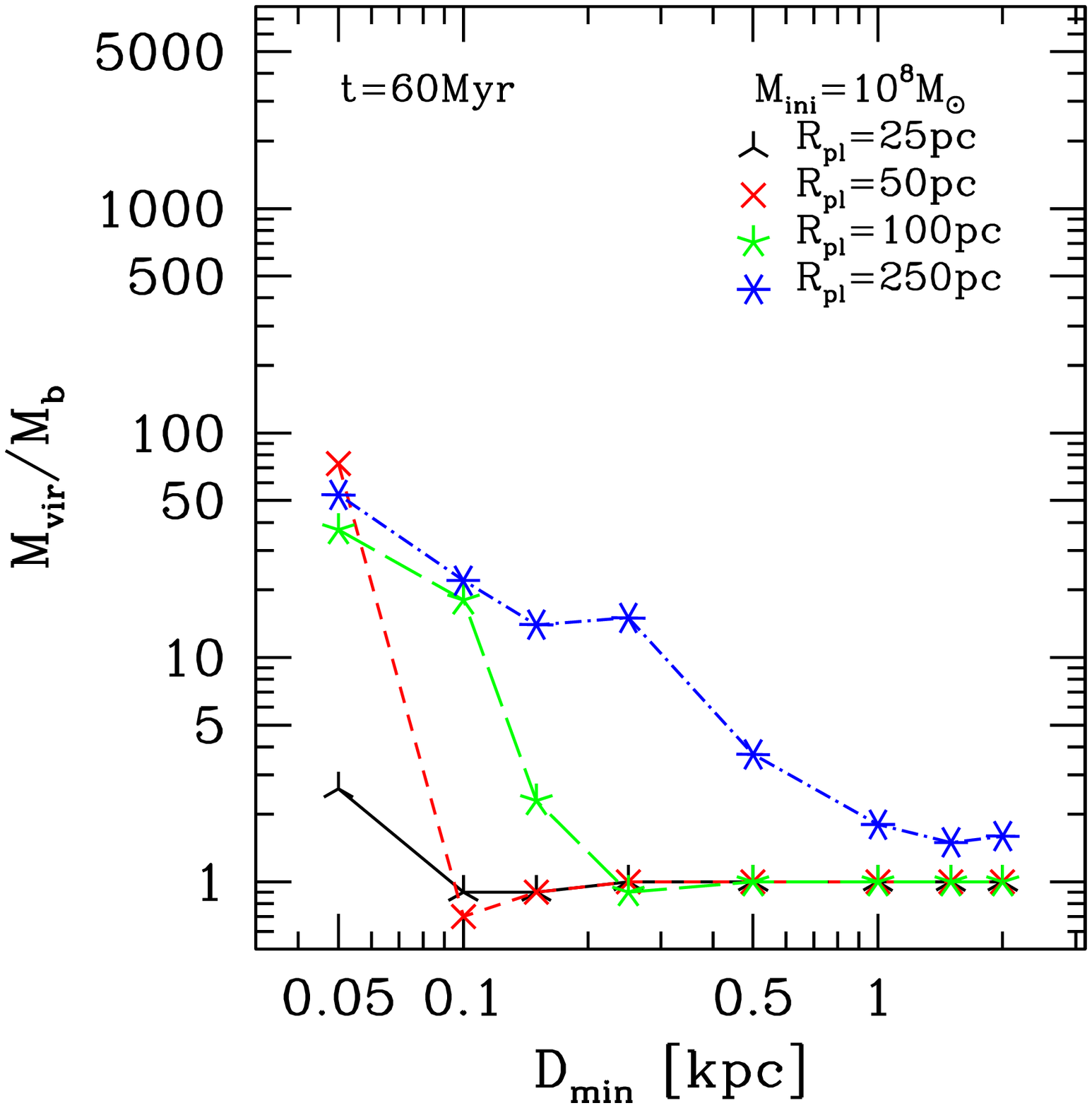}
    \caption{The ratio between virial and bound mass ($M_{\rm
        vir}/M_{\rm b}$) for our objects, measured at $t=60$~Myr, the
      point where the object is again most distant to the centre of
      the host galaxy, against the closest distance from the centre of
      the host ($D_{\rm min}$).  First panel: Objects with initial
      mass of $10^{7}$~M$_{\odot}$.  Second panel: Objects with
      initial mass of $10^{8}$~M$_{\odot}$. Tri-pointed stars
      (online: black) denote objects with initial Plummer radii of $R_{\rm
        pl}=25$~pc, crosses (online: red) have $R_{\rm pl}=50$~pc,
      five-pointed stars (online: green) have $R_{\rm pl}=100$~pc, and
      six-pointed stars (online: blue) $R_{\rm pl}=250$~pc.}
    \label{fig:all}
\end{figure}

\begin{figure}
    \centering 
    \epsfxsize=7cm 
    \epsfysize=7cm
    \epsffile{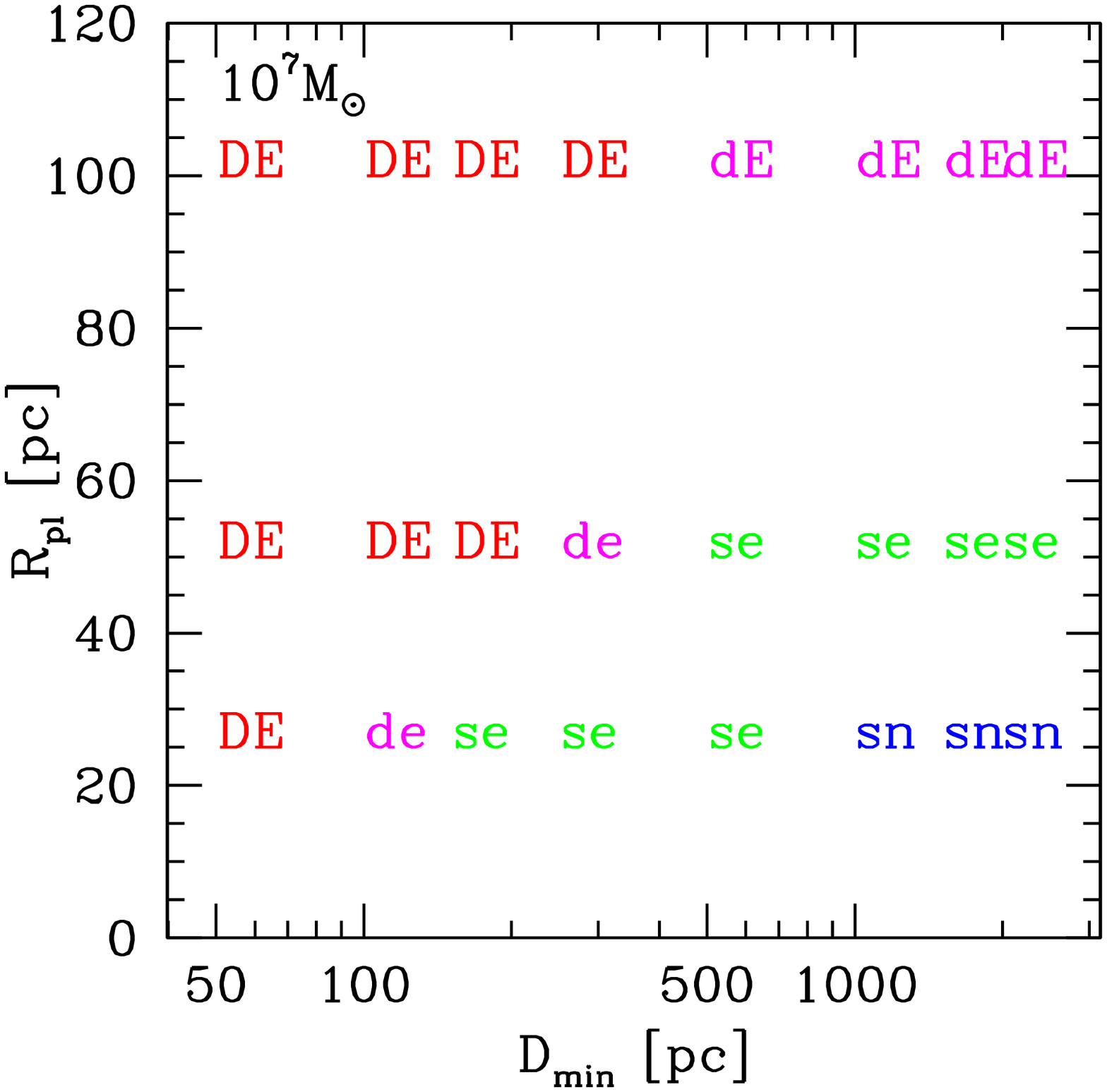}
    \epsfxsize=7cm 
    \epsfysize=7cm 
    \epsffile{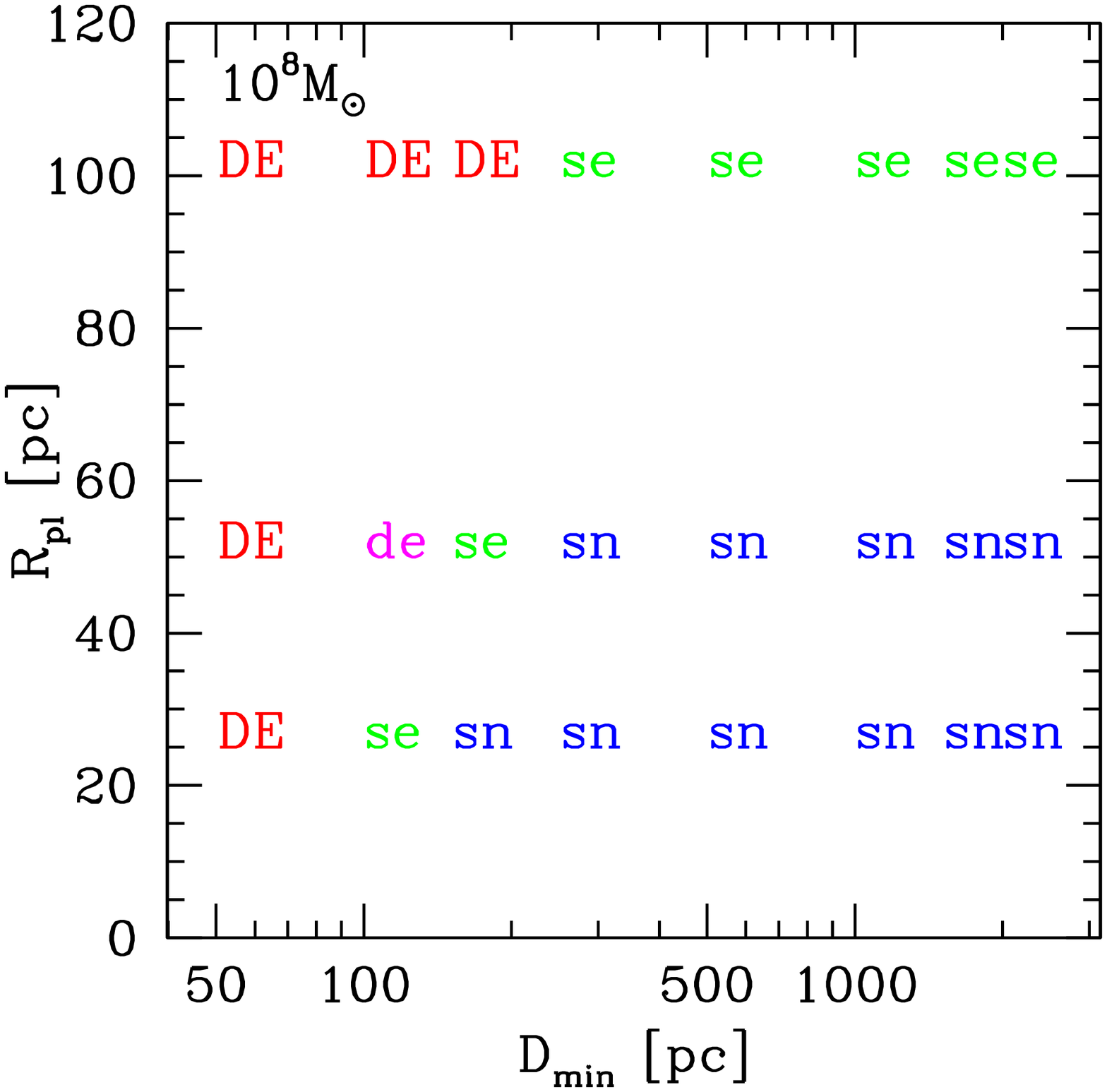}
    \caption{Parameter space of our simulations; Plotted are the
      Plummer radius of our models, $R_{\rm pl}$, against the closest
      distance to the centre of the host galaxy, $D_{\rm min}$.
      Capital D (online: red) denotes the satellite gets completely
      dissolved during the first passage.  Small d (online: magenta)
      denotes satellites which do not survive the next central
      passage.  Small s denotes satellites which survive and form
      bound objects even after the next central passage.  Capital E
      marks simulations where the M/M-ratio is enhanced if the central
      values are directly measured.  Small e marks simulations which
      only show enhanced M/M-ratios if the observational method is
      applied (see main text for explanation).  Surviving satellites
      with enhanced M/M-ratios are plotted green online.  Finally, n
      marks simulations which show no enhancement measured with either
      method (online: blue).  First panel is for objects with initial
      mass of $10^{7}$~M$_{\odot}$, second panel for
      $10^{8}$~M$_{\odot}$.} 
  \label{fig:result}
\end{figure}

We now determine the virial mass taking the formula from \citet{has05}
which is based on the theoretical work of \citet{kin66},
\begin{eqnarray}
  \label{eq:vir}
  M_{\rm vir} & = & \frac{9}{2 \pi G} \ \frac{\nu}{\alpha p} \ R_{\rm
    c} \ \sigma_{\rm 0,p}^{2},
\end{eqnarray}
where $R_{\rm c}$ is the core radius and $\sigma_{\rm 0,p}$ is the
central value of the projected velocity dispersion.  The parameters
$\nu$, $\alpha$, and $p$ are dependent on the kind of King model one
wants to fit.  Because we are using Plummer spheres we accumulate
these parameters together with the constants into one single parameter
$A$, which we determine for the isolated Plummer sphere before we
start the simulation,
\begin{eqnarray}
  \label{eq:virplum}
  M_{\rm vir} & = & A \ R_{\rm pl} \ \sigma_{\rm 0,p}^{2},
\end{eqnarray}
where the values of the scaling factor $A$ can be found in
Table~\ref{tab:para}.  As one can see these values are lower for
higher masses and increase for larger scale-lengths.  Nevertheless we
use the same $A$ determined for the isolated model for all final
models stemming from this initial model, even if the mass-loss is
significant.  This may lead to a slight underestimation of the final
virial mass.

\begin{figure*}
  \begin{minipage}{17.5cm}
    \centering 
    \epsfxsize=5.75cm 
    \epsfysize=5.75cm
    \epsffile{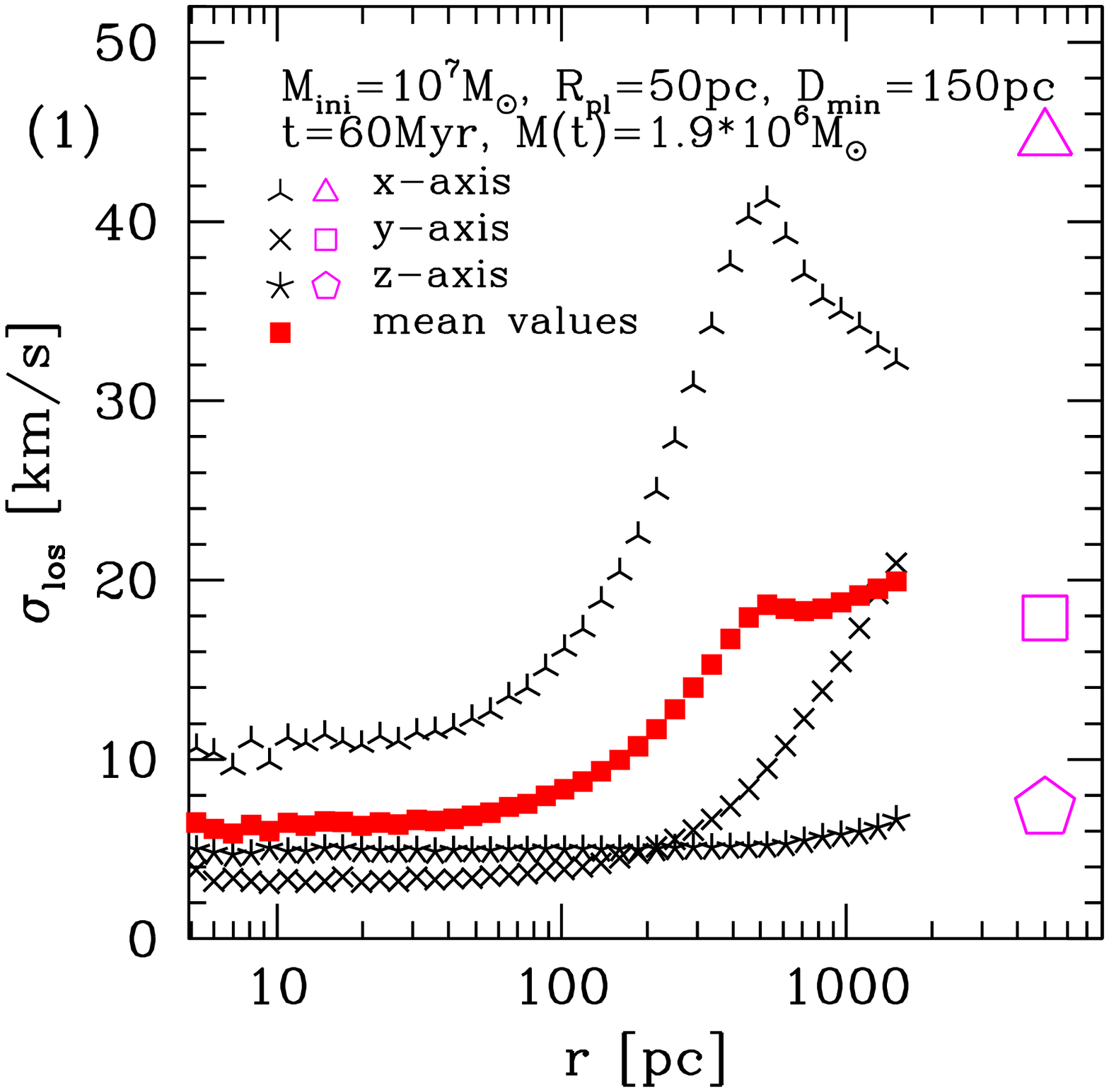}
    \epsfxsize=5.75cm 
    \epsfysize=5.75cm 
    \epsffile{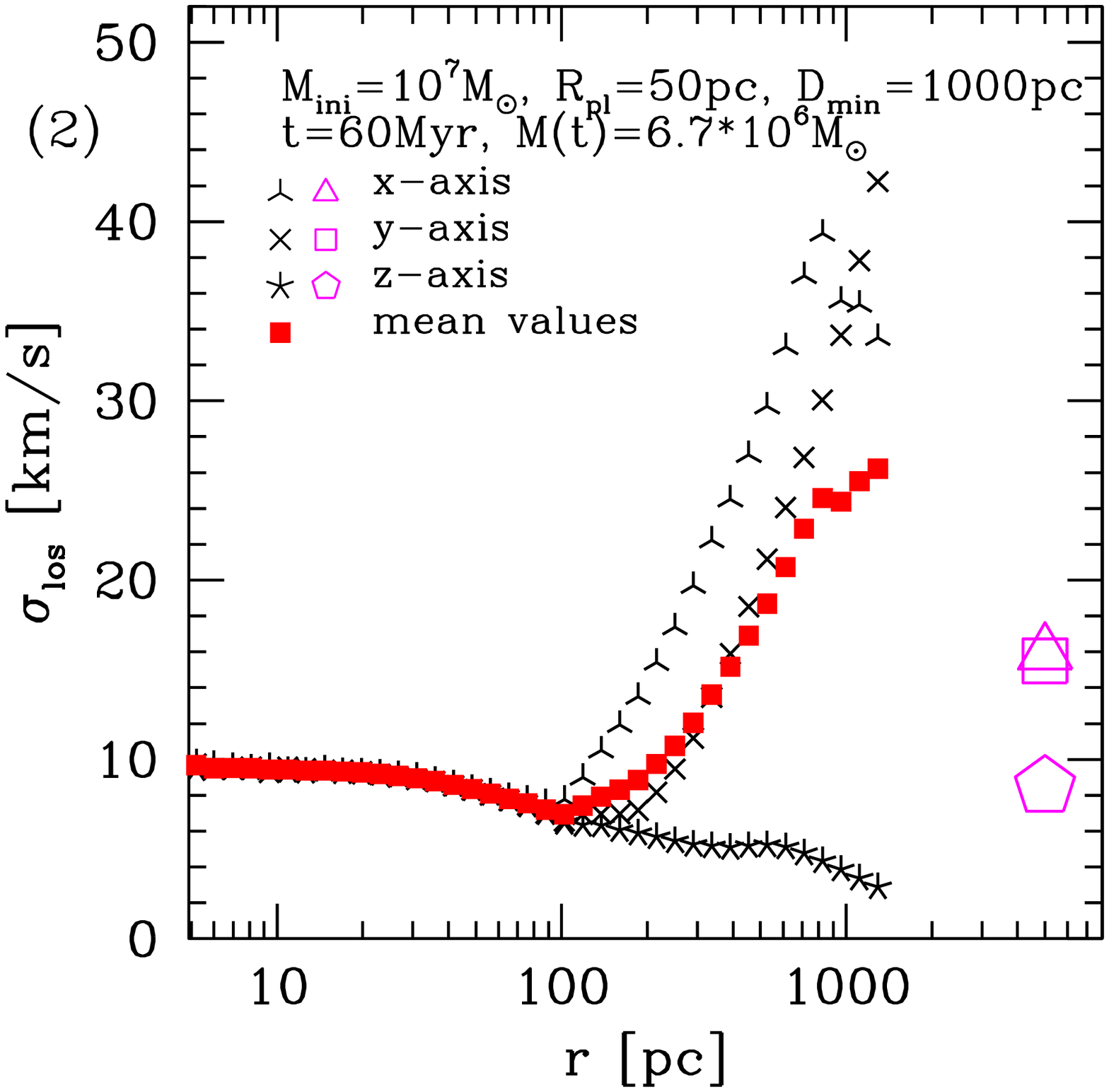}
    \epsfxsize=5.75cm 
    \epsfysize=5.75cm 
    \epsffile{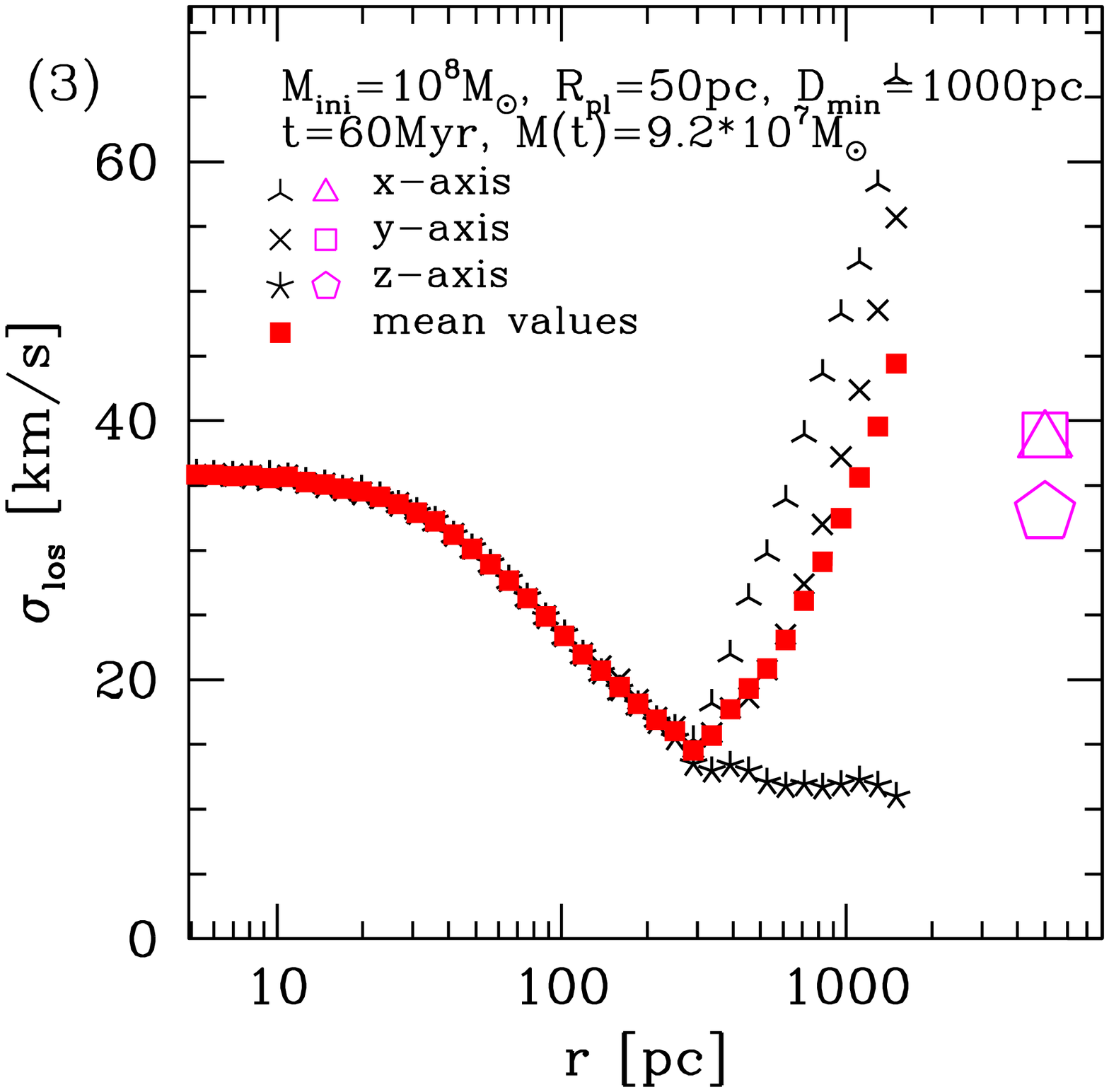}
    \caption{Line-of-sight velocity dispersion profiles and the
      Gaussian fit values for three of our satellites.  Three-, four-
      and five pointed stars denote the line-of-sight velocity
      dispersion values measured in concentric rings around the
      satellite along the $x$-, $y$ and $z$-axis, respectively.
      Filled squares (online: red) are the mean values of all
      directions.  Open symbols (online: magenta) on the right denote
      the central line-of-sight velocity dispersion values derived
      with the observational method (see Fig.~\ref{fig:gauss}).  First
      panel: This object gets completely dissolved and shows an
      enhancement of the velocity dispersion in either method.  But
      this destroyed satellite can not account for the UCDs in Virgo
      (see Fig.~\ref{fig:contour}).  Second panel: Object which shows
      an enhancement only if the central velocity dispersion is
      derived with the observational method.  The open symbols
      denoting the observational values in the $x$- and $y$-direction
      are clearly above the 'real' central value.  Third: Simulation
      of a very massive satellite which shows no clear enhancement in
      both methods.}
    \label{fig:rmax}
  \end{minipage}
\end{figure*}

\begin{figure*}
  \begin{minipage}{17.5cm}
    \centering 
    \epsfxsize=5.75cm 
    \epsfysize=5.75cm
    \epsffile{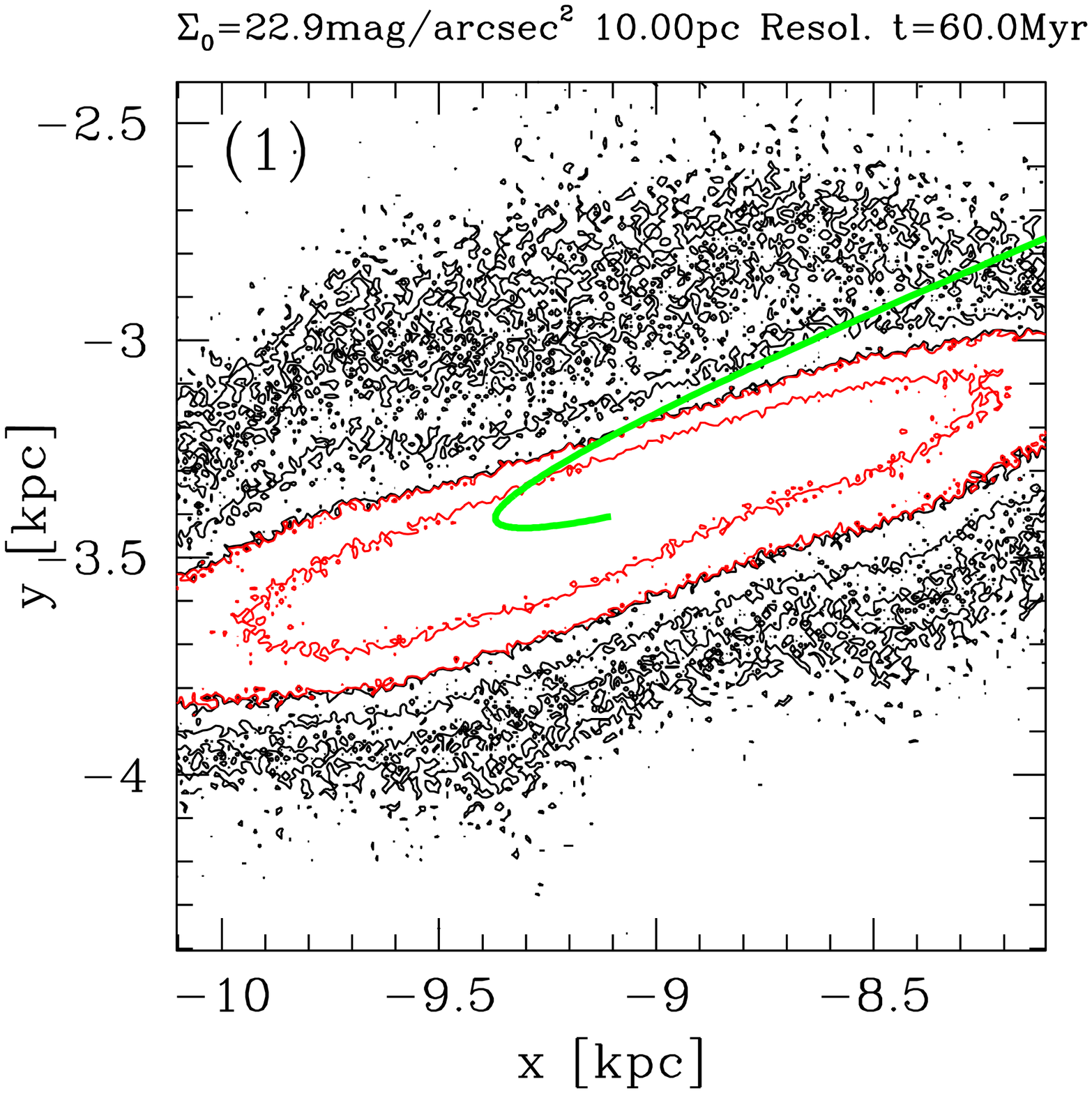}
    \epsfxsize=5.75cm 
    \epsfysize=5.75cm 
    \epsffile{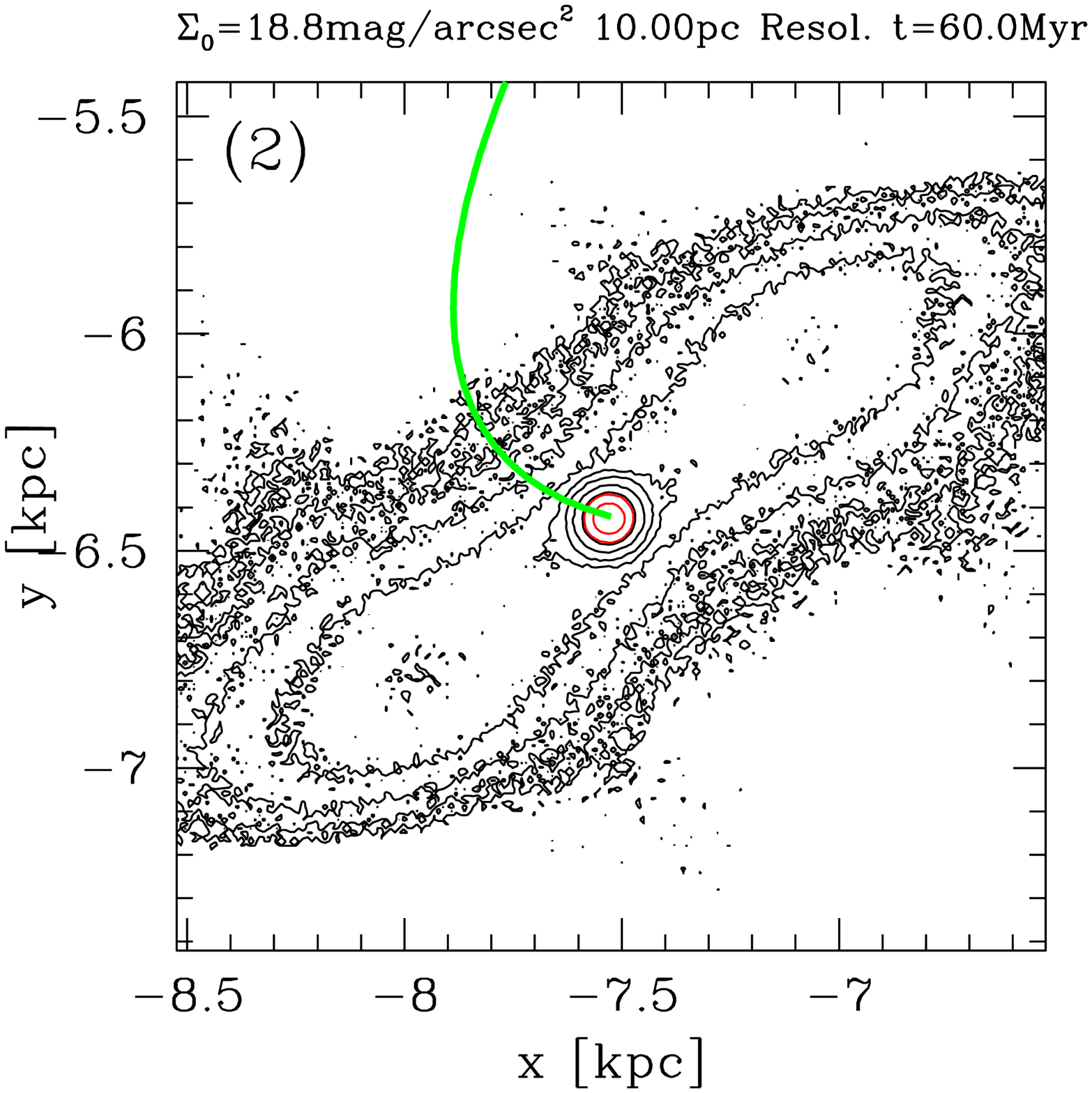}
    \epsfxsize=5.75cm 
    \epsfysize=5.75cm 
    \epsffile{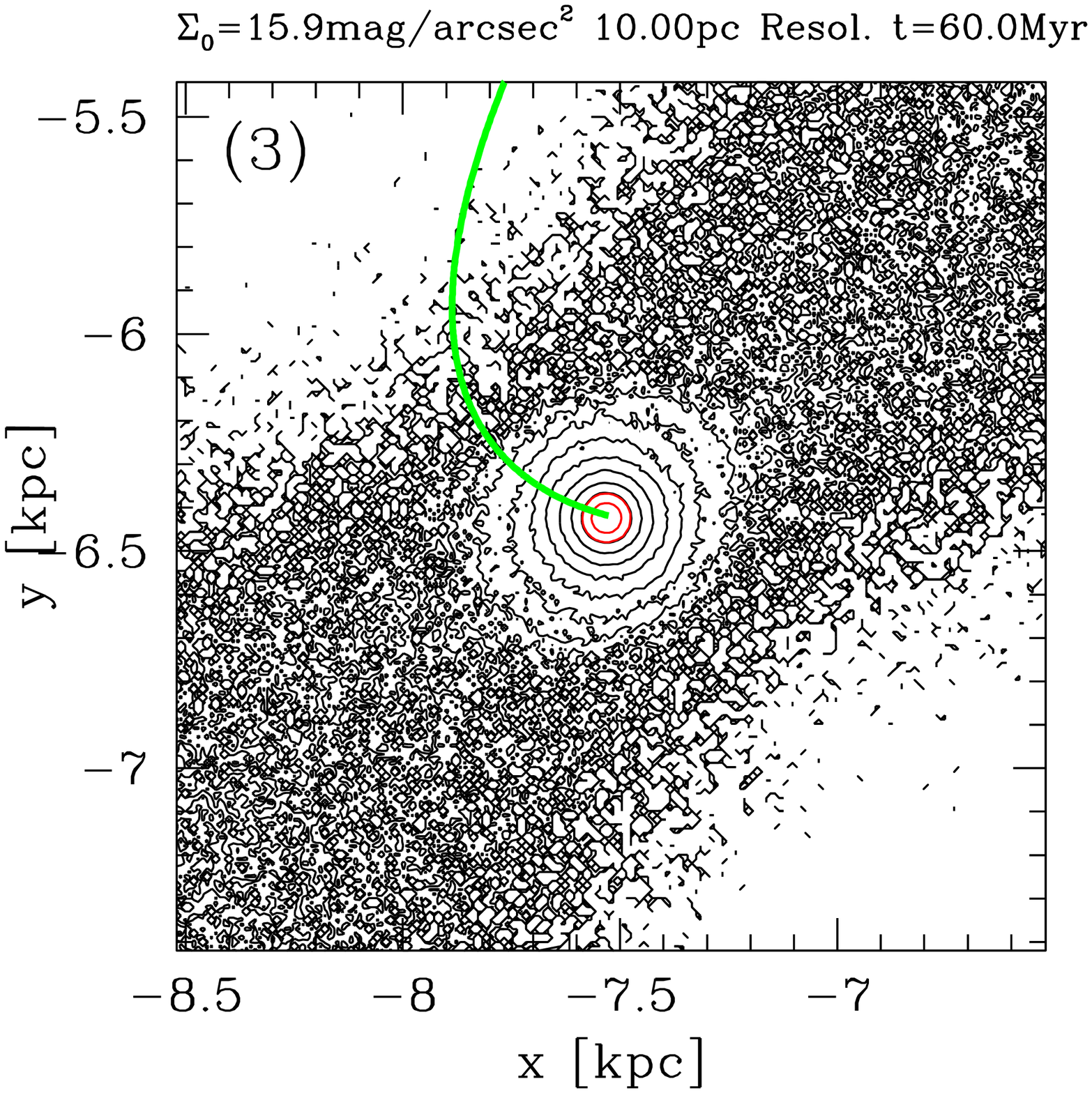}
    \caption{Contour plots of the three models of Fig.~\ref{fig:rmax}
      in the orbital plane ($x$-$y$-plane).  The centre of the host
      galaxy is at $0$--$0$~kpc.  The orbit of the object is plotted
      as the thick (online: green) line.  Contours have magnitudo spacing
      (assumed $M/L = 1.0$~M$_{\odot}$/L$_{\odot}$) and the outermost
      contour corresponds to $28$~mag\,arcsec$^{-2}$ (the two brightest
      contours in each panel are plotted red online).  Almost filled
      black areas do not denote high surface brightness but areas with
      very faint contributions (at the given resolution the brightness
      fluctuates from pixel to pixel at the lowest level).  First
      panel: This object is clearly dissolved; it shows no core
      anymore and is out of virial equilibrium.  Second: Surviving
      object which shows faint tidal tails along the orbit which
      enhance the measured velocity dispersion.  Third: This object
      shows only a few stars which are unbound and spread (black
      areas).  The small fraction of these stars are not able to
      enhance the velocity dispersion.}
    \label{fig:contour}
  \end{minipage}
\end{figure*}

\begin{figure*}
  \begin{minipage}{17.5cm}
    \centering 
    \epsfxsize=5.75cm 
    \epsfysize=5.75cm
    \epsffile{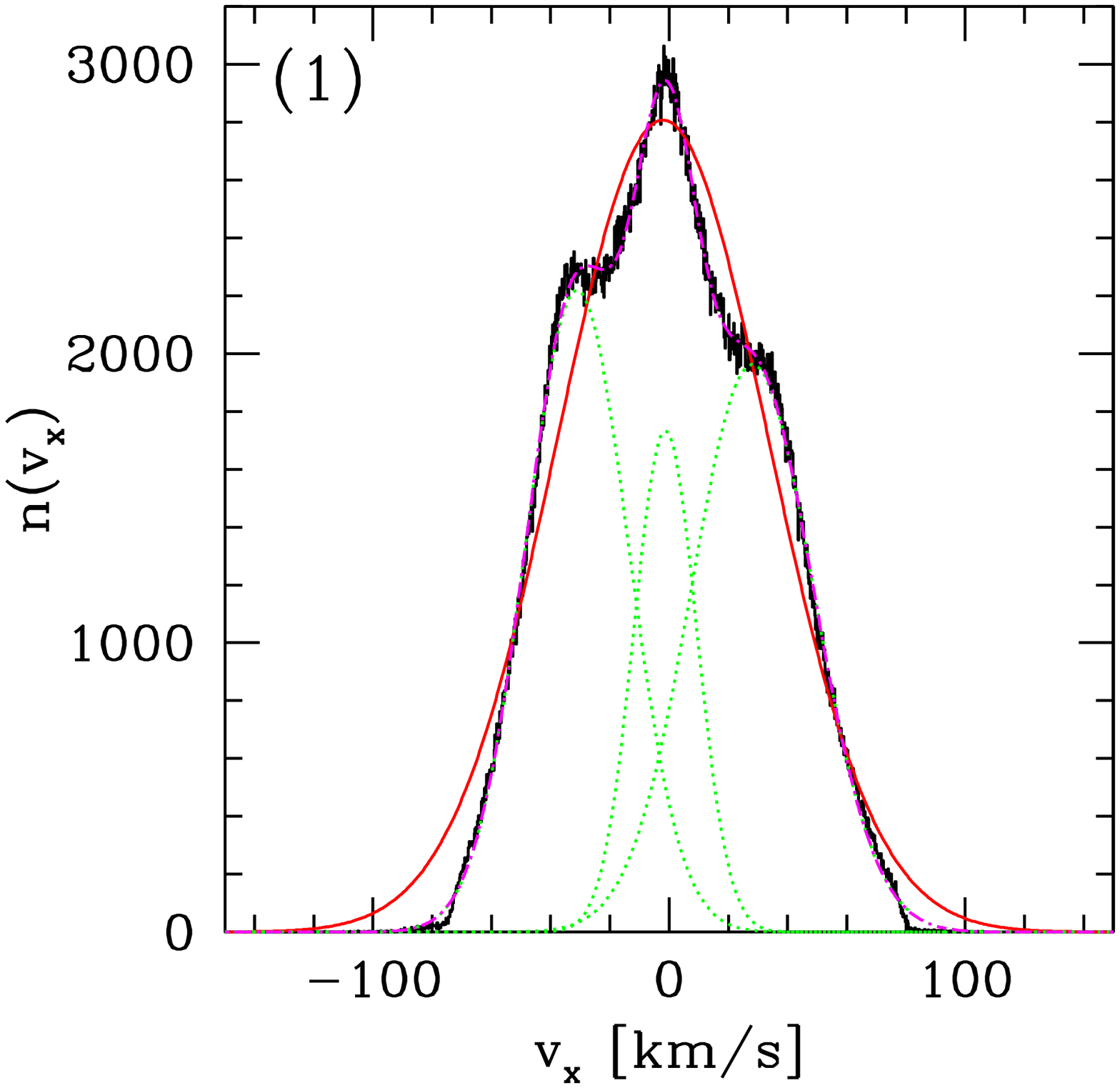}
    \epsfxsize=5.75cm 
    \epsfysize=5.75cm 
    \epsffile{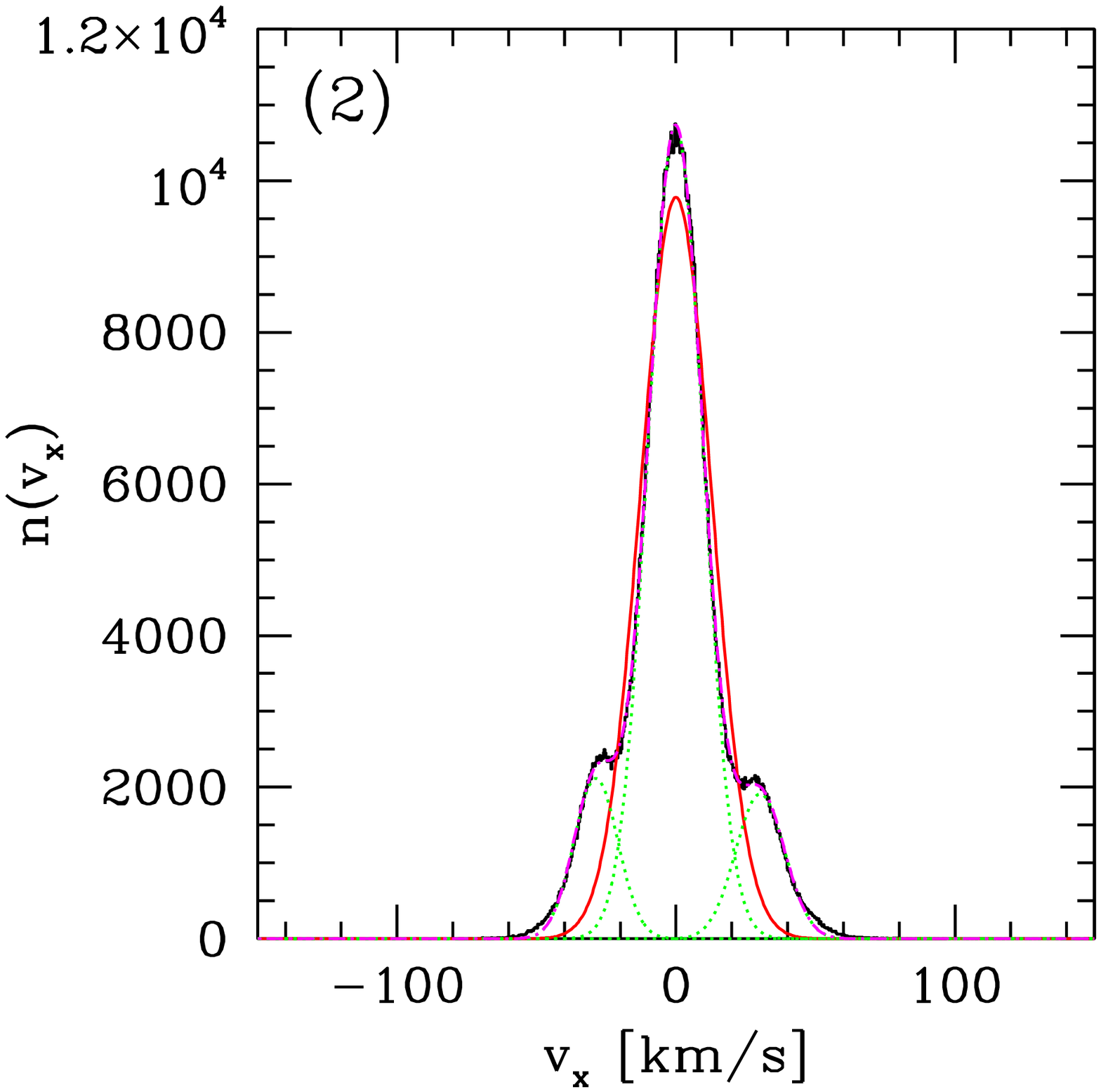}
    \epsfxsize=5.75cm 
    \epsfysize=5.75cm 
    \epsffile{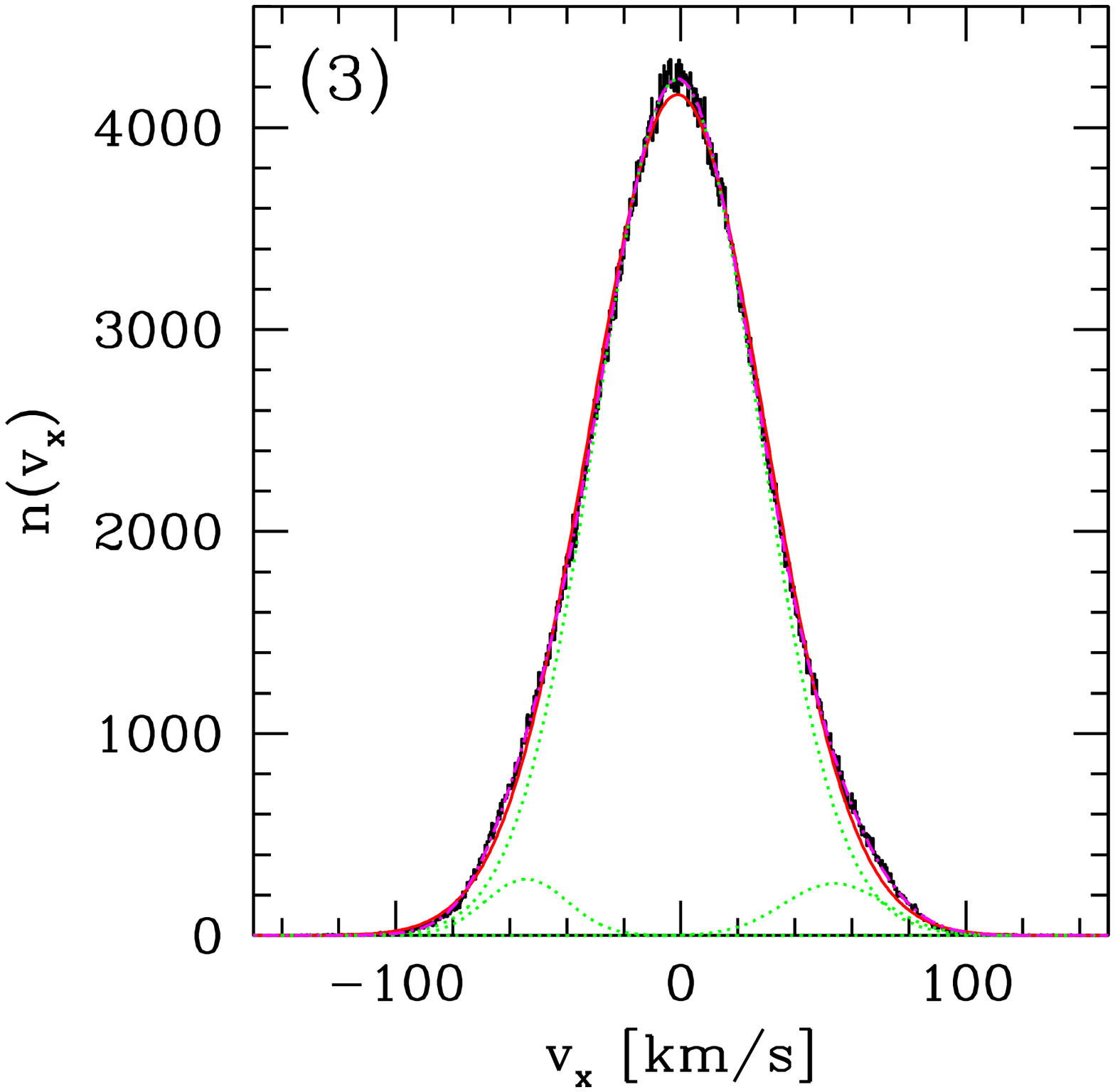}
    \caption{Same simulations as in Fig.~\ref{fig:rmax}.  The plots
      show the velocity distributions along the x-axis.  Histogram
      (online: black) are the data-points.  Solid line (online: red) is
      the single Gaussian fit, dotted lines (online: green) are the
      three Gaussians to correct for fast and slow unbound particles
      in front and behind the object and the dashed line (online: 
      magenta) is the sum of these three Gaussians, which is an
      excellent reproduction of the data and is barely visible by
      being masked by the data.  The three Gaussians always fit the
      data best.  While in the first panel the object is almost
      dissolved and the Gaussians of the unbound particles are
      stronger than the bound one (resulting in a M/M-ratio of $20$ in
      $x$ and $5$ in $y$-direction), the second panel shows a bound
      object with enhanced M/M-ratio due to the small contribution of
      the unbound particles (M/M-ratio is $2.5$ and $2.2$).  In the
      last panel the contribution of the unbound particles is too low
      to enhance the M/M-ratio (M/M $=1.2$ in both directions and
      within the uncertainties of the measurement).}
    \label{fig:gauss}
  \end{minipage}
\end{figure*}

On the other hand {\sc Superbox} calculates the number of bound
particles (energy below zero) at each time-step.  Therefore we know
the real bound mass $M_{\rm b}$.

With these two masses we can determine a virial-mass-to-bound-mass
ratio ($M_{\rm vir} / M_{\rm b}$ or short M/M) which can then be
multiplied by a 'normal' mass-to-light ratio for a population of stars
with the determined age and metallicity and without dark matter to
obtain the dynamically measured $M/L$-ratio.

This M/M-ratio is plotted in Fig.~\ref{fig:all}, when the satellites
have reached their apogalactica again.  At this point the satellites
are almost back to virial equilibrium again if they are not on the way
to complete dissolution.

The results show clearly that only objects which are not compact
and/or have a low mass can be influenced enough by one central passage
to enhance the M/M to account for the high mass-to-light ratios found
in UCDs.  But these satellites will also not survive the next central
passage or have not survived the first one at all.

The best representation of real UCDs is the model with an initial mass
of $10^{7}$~M$_{\odot}$ and an initial Plummer radius of $25$~pc.  The
results for this model are shown as tri-pointed stars in the first
panel of Fig.~\ref{fig:all} and the lowest line in the first panel of
Fig.~\ref{fig:result}.  The results show clearly that either the
satellite gets completely dissolved if $D_{\rm min}$ is closer than
$50$--$100$~pc or shows no deviation in M/M at all
(Fig.~\ref{fig:result}). 

In Fig.~\ref{fig:result} we distinguish the models according to their
survival of the passage and if they show enhanced M/M-ratios.  The
panels show clearly that only dissolved or almost dissolved objects
show enhanced M/M-ratios.  All surviving objects show no deviation
from virial equilibrium.

\subsection{The 'observational' method}
\label{sec:obs}

The results in Sect.~\ref{sec:theo} are based on the exact knowledge
of the theoretical central line-of-sight velocity dispersion.
Observers on the other hand do not have this information.  The usual
way to determine the velocity dispersion of a marginally resolved
object is to place a slit on the object and obtain a spectrum.  From
this spectral information one chooses one or two spectral lines and
fits a template spectrum convolved with the instrumental line-width
function and a Gaussian velocity distribution.  This Gaussian velocity
distribution measures the line-of-sight velocity dispersion of the
whole object.  Based on assumed theoretical models this value is then
corrected to the central line-of-sight velocity dispersion.  In the
case of the simple Plummer model one has to multiply the result for
the whole object by a factor of 1.25.  Thus, the central velocity
dispersion is  
\begin{eqnarray}
  \label{eq:gauss1}
  \sigma_{0,p}^{\rm pl} & = & \sqrt{\frac{3 \pi G M_{\rm pl}} {64
      R_{\rm pl}}},
\end{eqnarray}
while the projected velocity dispersion integrated over the whole
Plummer sphere is
\begin{eqnarray}
  \label{eq:gauss2}
  \sigma_{\rm obs,p}^{\rm pl} & = & \frac{1}{M_{\rm pl}} \
  \int_{0}^{\infty} 2 \pi r' \Sigma(r') \sigma_{p}(r') {\rm d}r',
  \nonumber \\
  & = & \sqrt{\frac{3 \pi G M_{\rm pl}} {100 R_{\rm pl}}},
\end{eqnarray}
so that
\begin{eqnarray}
  \label{eq:gauss3}
  \frac{\sigma_{0,p}^{\rm pl}} {\sigma_{\rm obs,p}^{\rm pl}} & = & 1.25.
\end{eqnarray}
$\sigma_{0,p}^{\rm pl}$ denotes the central line-of-sight velocity
dispersion and $\sigma_{\rm obs,p}^{\rm pl}$ the weighted
line-of-sight velocity dispersion integrated over the whole object.

\begin{figure*}
  \begin{minipage}{17.5cm}
    \centering 
    \epsfxsize=5.75cm 
    \epsfysize=5.75cm
    \epsffile{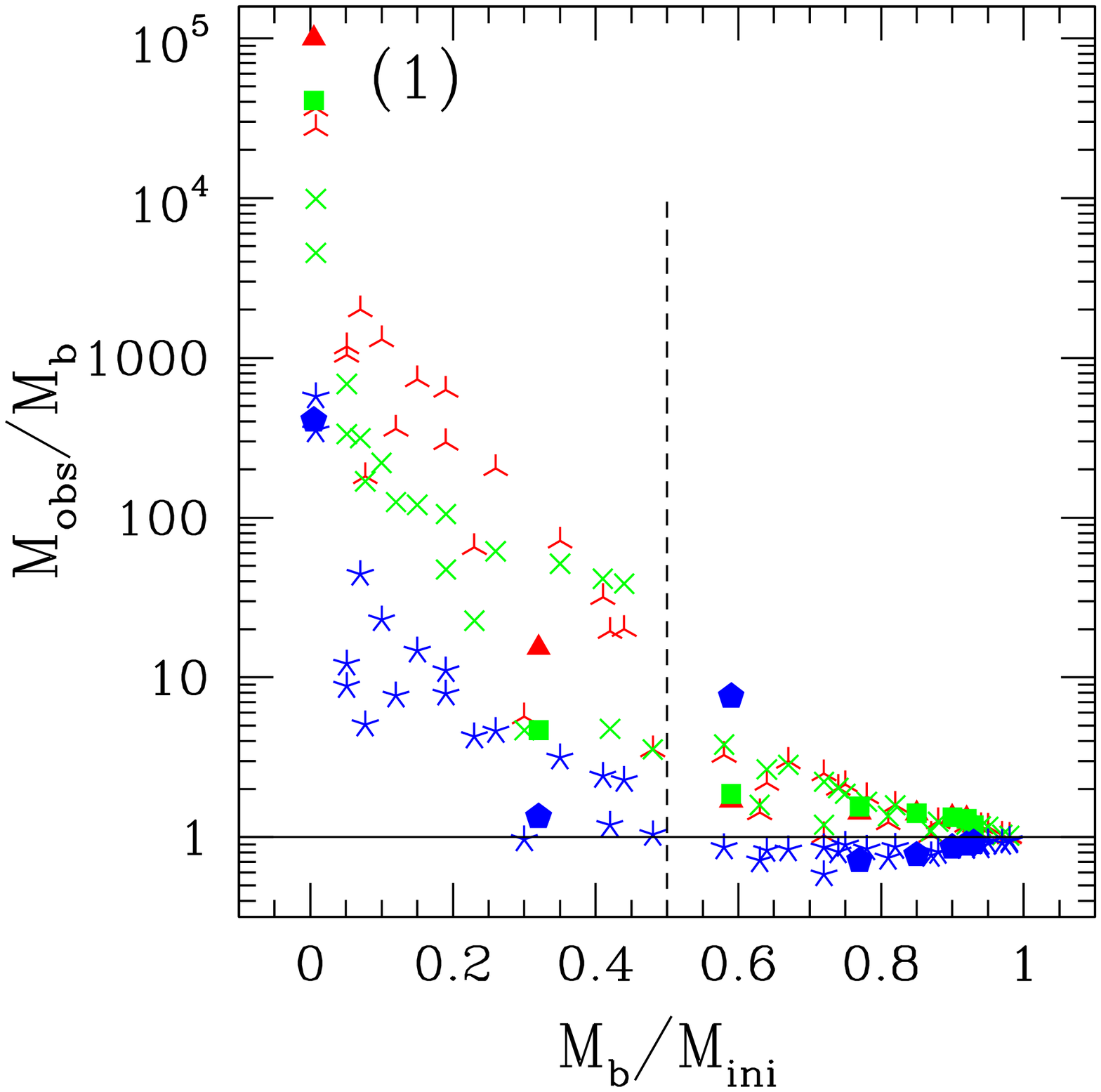}
    \epsfxsize=5.75cm 
    \epsfysize=5.75cm 
    \epsffile{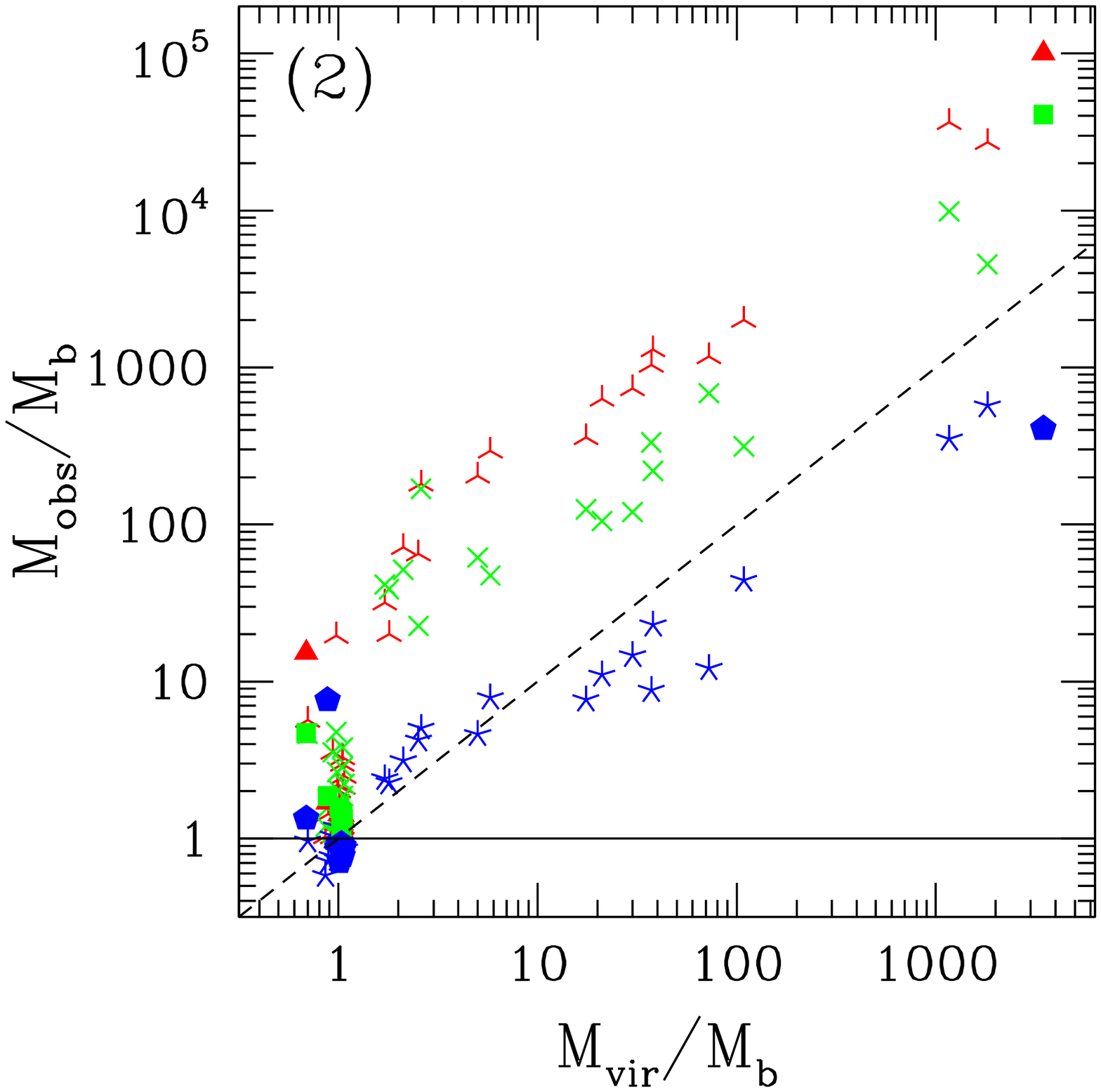}
    \epsfxsize=5.75cm 
    \epsfysize=5.75cm 
    \epsffile{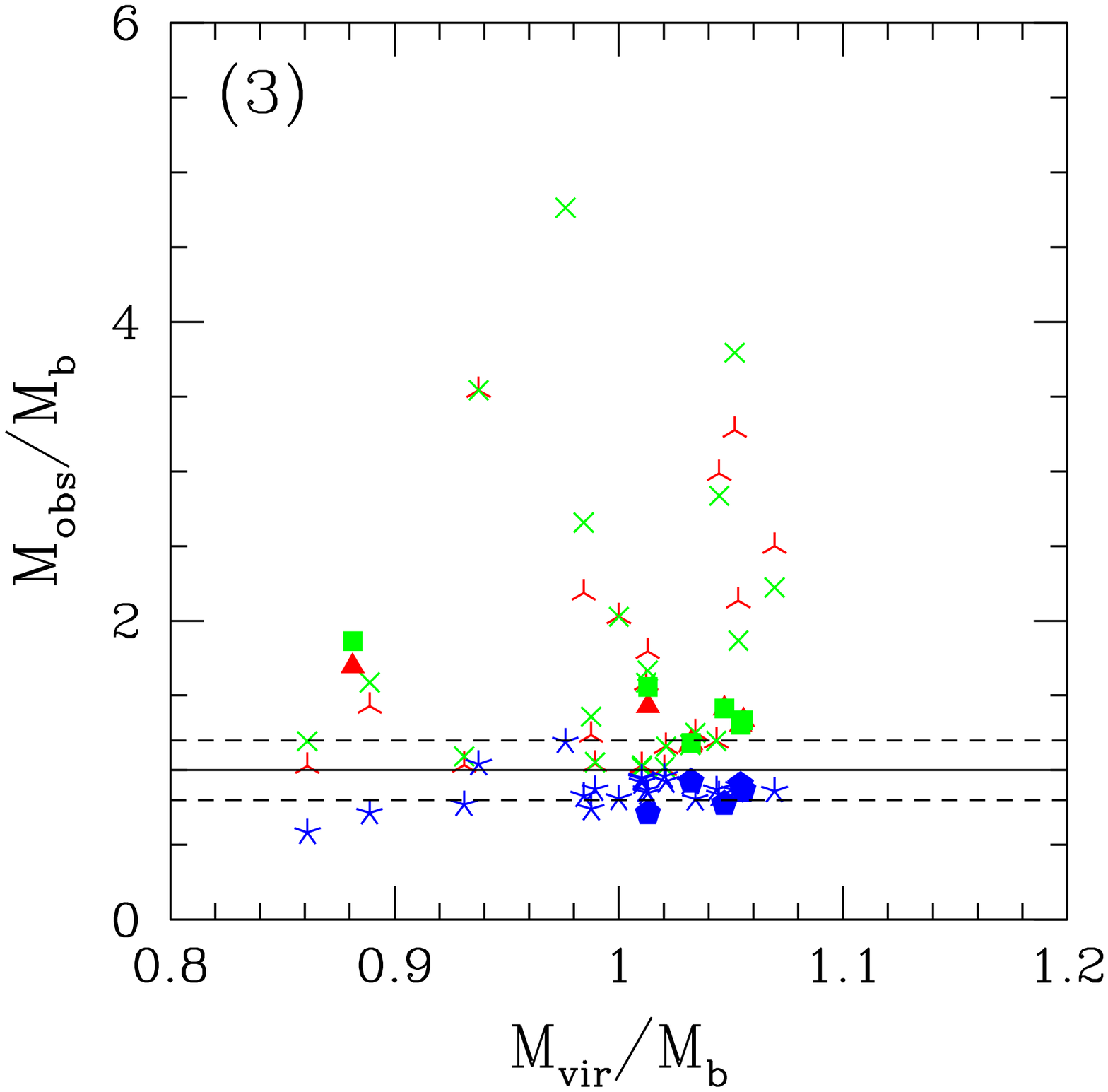}
    \caption{Summary of the results derived using the observational
      method (Sect.~\ref{sec:obs}).  $M_{\rm obs}$ denotes the virial
      mass derived by the observational method, i.e. fitting a single
      Gaussian to the velocity distribution, $M_{\rm b}$ denotes the
      real bound mass of the object, $M_{\rm ini}$ is the initial
      bound mass and $M_{\rm vir}$ is the virial mass derived by the
      analytical method (Sect.~\ref{sec:theo}).  Measurements along
      the $x$-axis are three-pointed stars (online: red), along the
      $y$-axis are crosses (online: green) and $z$-axis measurements
      are represented by five-pointed stars (online: blue).
      Simulations with initial parameters similar to the UCDs in Virgo
      are represented by filled symbols.  First panel shows the
      derived mass-ratio as a function of the final/initial
      bound-mass-ratio.  Satellites with final masses larger than 50
      per cent of the initial mass (right of the dashed line) are
      deemed stable against immediate tidal destruction.  Objects left
      of the dividing line are out of equilibrium and therefore have
      to show enhanced M/M-ratios.  But there are also data points
      which are larger than one on the right of the dividing line.
      Second panel shows the same results but now the observational
      M/M-ratio is plotted against the analytical ratio.  All
      data-points except the measurements perpendicular to the orbit
      are above the 'analytical' method (dashed line shows the line
      where the values of the observational method are equal to the
      direct method).  $z-$values are below the dashed line because we
      do not correct the constant $A$ for objects with high mass-loss.
      A clump of data-points is visible around $M_{\rm vir}/M_{\rm
        b}=1$ which points to overestimated masses.  The third panel
      finally is a magnification of this region.  One clearly sees
      that the observational method can lead to overestimates of the
      true mass by a factor of up to five.  Horizontal line shows a
      ratio of $1$ with dashed lines denoting an error range of
      $20$~per cent.}
    \label{fig:obs}
  \end{minipage}
\end{figure*}

As one can see in Fig.~\ref{fig:rmax} it can be a crucial point how
strongly a measurement of the line-of-sight velocity dispersion is
contaminated by unbound stars.  The observational values can highly
overestimate the real central value.  Unbound stars have a different
velocity distribution than the bound ones.  They are either travelling
in front or behind the object (seen along the trajectory; this is
shown in Fig.~\ref{fig:contour}) and are faster or slower.  But the
velocity distribution of these stars which are still located in the
vicinity of the object and stem from one central passage, peak on
either side of the Gaussian distribution of the bound stars, leading
to an effective broadening of the measured velocity distribution.
Fitting just a single Gaussian will therefore lead to an enhanced
measured velocity dispersion and an overestimation of the central
velocity dispersion value of the satellite.  This is demonstrated in
Fig.~\ref{fig:gauss} where the velocity distribution of all stars is
shown together with the best fitting single Gaussian and the fitting
curve of a triple Gaussian which takes the unbound stars into account.
The measured values for the central line-of-sight velocity dispersion
are shown in Table~\ref{tab:sig}.

\begin{table}
  \centering
  \caption{Central velocity dispersion along the x-axis derived by
    different methods.  Shown are the values of the same three
    simulations plotted in Figs.~\ref{fig:rmax} to~\ref{fig:gauss}
    now labelled 'dissolved', 'enhanced' and 'massive', respectively.
    The rows show 
    the different values of the central velocity dispersion.  First
    row is the direct measurement of the central dispersion, 'single'
    denotes the central dispersion derived from the fit of a single
    Gaussian to the data and 'triple (c)' denotes the value of a fit
    using three Gaussians where the central one (c) is used to
    compute the velocity dispersion of the bound object.}
  \label{tab:sig}
  \begin{tabular}[t!]{r|rrr} \hline
    [km/s] & dissolved & enhanced & massive \\ \hline \hline
    true actual & $10.6 \pm 3.1$ & $9.49 \pm 0.05$ & $35.7 \pm 0.2$
    \\  
    single   & $44.6 \pm 0.3$ & $15.90 \pm 0.30$ & $38.4 \pm 0.1$ \\
    triple (c) & $13.3 \pm 0.1$ & $12.35 \pm 0.04$ & $35.3 \pm 0.3$
    \\ \hline
  \end{tabular}
\end{table}

Clearly this effect is strongest in the plane of the orbit and is not
visible perpendicular to it.  This can be seen in Fig.~\ref{fig:rmax}
where the symbol for the $z-$axis-value in all three panels shows the
same value as the direct measurement.  Actually for strongly disturbed
systems (i.e.\ high mass-loss) the values in the $z$-direction are
below the line because we do not account for the change of the
constant $A$ which should increase for lower masses.  But in any
random orientation of the object with respect to the observer the
effect of the enhanced M/M-ratio should at least be partly visible.

Summing up our results we state the following: Satellites which are
out of virial equilibrium, i.e.\ in the state of dissolution show
enhanced virial masses and therefore the real mass content is
overestimated.  Figure~\ref{fig:obs} shows in the first panel the
M/M-ratio plotted against the ratio of final to initial bound mass of
the object.  The dividing dashed line separates objects which have
lost more than $50$~per cent of their mass during the central passage
and are already dissolved or are not likely to survive the next
passage, from objects which are stable for several more passages
through the centre.  While on the left side the derived M/M-ratios
(using the observational method) can climb up to very high values, the
stable objects show only slight enhancements of the ratio if at all.
But in the second panel one already sees that even if the object
itself is in virial equilibrium ($M_{\rm vir}/M_{\rm b} \approx 1.0$)
there are satellites which show a broadening of the velocity
distribution leading to an enhanced mass-ratio if measured the
observational way.  The third panel finally shows an enlargement of
this area and one finds M/M-ratios overestimated by up to a factor
of five.  Surviving objects which show an enhancement in their
M/M-ratio, if measured with this observational method, are already
marked in Fig.~\ref{fig:result} with a small 'e'.  As one can see
there is a range of critical distances ($D>100$~pc and $D<1$~kpc) to
the centre of the host galaxy where an UCD like the ones found in
Virgo could show an overestimated mass-to-light ratio.

\subsection{Observability}
\label{sec:observ}

In the previous section we claimed that with the 'observational'
method the mass-to-light ratios of UCDs could be overestimated,
because the velocity distribution is not Gaussian any more but
'contaminated' by unbound stars around the object.  In this section we
show that there is almost no chance for an observer to detect
this 'non-Gaussianity' of the velocity distribution.  

When measuring a spectrum of a distant object, observers have to deal
with two major shortcomings.  First there is the intrinsic line-width
produced mainly by the instrument.  State-of-the-art instruments like
UVES or Flames can reduce this line-width down to about
$2$~km\,s$^{-1}$.  But the observations of the UCDs in Virgo were made
with an instrumental line-width of about $25$~km\,s$^{-1}$.

The second effect an observer has to take into account is noise in the
spectrum.  

\begin{figure}
  \centering
  \epsfxsize=9cm
  \epsfysize=13.5cm
  \epsffile{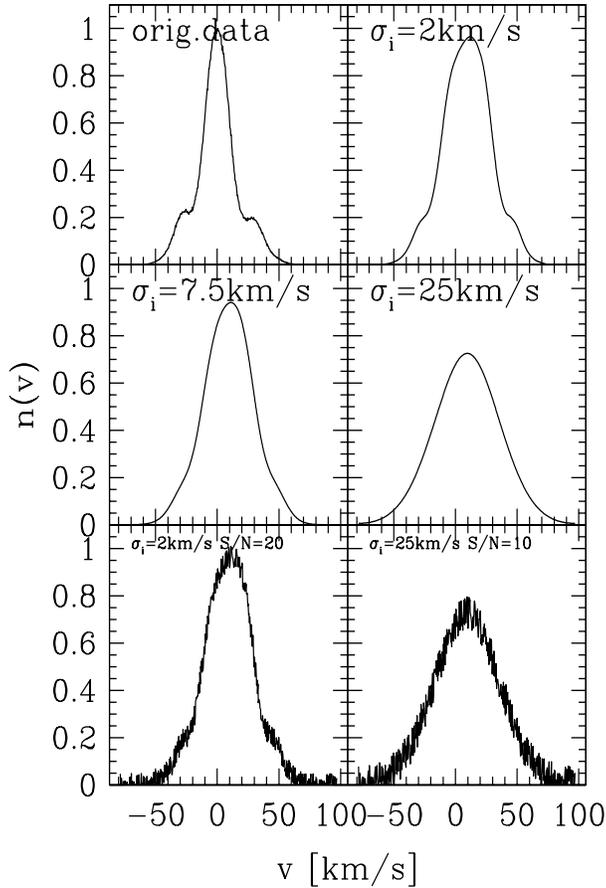}
  \caption{The top left panel shows the original velocity
    distribution.  At the top-right panel this distribution was folded
    with a Gaussian of the width $2$~km\,s$^{-1}$ which mimics the
    best intrinsic instrumental line-width available today.  The
    middle panels show the same distribution but folded with a
    Gaussian of the width $7.5$~km\,s$^{-1}$ (left) and
    $25$~km\,s$^{-1}$ (right).  The left panel shows the transition
    when the non-Gaussian features of the original distribution start
    to disappear.  In the bottom panels we also added white noise to
    the folded distribution to mimic the noise an observer has in the
    spectrum.  At the left panel we took the folded distribution with
    the best available line-width again and added noise to reach a S/N
    of $20$.  This is again the transition where the features start to
    disappear.  With more noise the features are not visible any
    more.  Clearly with higher intrinsic line-widths the features
    disappear at lower noise-levels.  The right panel mimics the
    properties of the spectra of the UCDs in Virgo \citep{has05}.
    There the intrinsic line-width was $25$~km\,s$^{-1}$ and the
    S/N-ratio was $10$.}
  \label{fig:dist}
\end{figure}

Taking the velocity distribution from the enhanced M/M-ratio
simulation we fold it with a Gaussian of the width $\sigma_{\rm i}$ to
mimic the instrumental line-width and determine at which line-width
the 'features' of our distribution are washed out.  This happens at an
instrumental line-width of $\sigma_{\rm i}=7.5$~km\,s$^{-1}$ (as shown
in Fig.~\ref{fig:dist}).  Then we take the best state-of-the-art
line-width of $\sigma_{\rm i}=2$~km\,s$^{-1}$ and add random white
noise to the distribution until again the 'features' are almost
invisible again.  This happens already at a signal-to-noise ratio of
$20$ (see Fig.~\ref{fig:dist} lower left panel).  In the final panel
of Fig.~\ref{fig:dist} we fold our distribution with the line-width of
the observations of the Virgo-UCDs and add noise to mimic the same
S/N-ratio as in the observations.  There is no deviation from
Gaussiantity visible any more.

\section{Discussion \& Conclusion}
\label{sec:disc}

We have shown with our models that dwarf satellites around a giant
elliptical galaxy like M87 can have an enhanced mass-to-light ratio
due to close passages to the centre of the host galaxy.  While very
close passages lead to the destruction of the satellite there are
orbits which allow for enough 'damage' to the satellite to enhance the
mass-ratio (measured the same way an observer would do) without
completely destroying the object.

The amount of destruction is larger if the satellite is less massive
and less concentrated.  But the loss of about $20$~per cent of the
initial mass is enough to have the object surviving several more close
passages and to mimic an enhanced mass-to-light ratio.  For models
comparable to the UCDs in Virgo and Fornax the range of possible
minimum distances during central passages is about $100$ to $1000$~pc.
All passages closer to the centre lead to complete destruction and all
passages further away show no measurable effect at all, except for a
mass-loss of the order of a few per cent.  We therefore conclude that
the enhanced M/L-ratios measured for the Virgo UCDs by \citet{has05}
may not be due to dark matter.

A more general result of our study is the discrepancy between the
derived virial masses if one has access to the correct properties of
the satellites compared to the virial masses derived the way an
observer would measure.  While very massive objects which are almost
unaffected by tidal forces show the same results within the
uncertainties one has to be careful if objects are less massive and
are surrounded by a cloud of tidally stripped stars.  These stars are
either faster or slower in the mean but their mean values are not too
different to the bulk velocity of the bound stars to clearly
disentangle the 'populations' (populations in the sense of within, in
front or behind the object).  Especially if the broadening of a
spectral line is estimated by fitting the template line folded with
a Gaussian for the instrumental line-width and a single Gaussian for
the velocity distribution can the deduced central velocity dispersion
be too high thus leading to a mass-to-light ratio that is too large by
up to a factor of ten.

Effects like this have to be given serious consideration when
measuring velocity dispersions of faint and distant objects. \\

\noindent {\bf Acknowledgements:}

\noindent MF thankfully announces financial support through DFG-grant
KR1635/5-1 and PPARC.  We also want to thank M. Hilker and T. Richtler
for useful comments regarding how to mimic observations.

\label{lastpage}


\begin{thebibliography}{}

\bibitem[\protect\citeauthoryear{Aarseth, Henon \& Wielen}{Aarseth et
    al.}{1974}]{aar74} 
  Aarseth S.J., Henon M., Wielen R., 1974, A\&A, 37, 183

\bibitem[\protect\citeauthoryear{Bekki, Couch \& Drinkwater}{Bekki et
    al.}{2001}]{bek01} 
  Bekki K., Couch W.J., Drinkwater M.J., 2001, ApJ, 552, 105L

\bibitem[\protect\citeauthoryear{Bekki et al.}{2003}]{bek03} 
  Bekki K., Couch W.J., Drinkwater M.J., Shioya Y., 2003, MNRAS, 344,
  399 

\bibitem[\protect\citeauthoryear{Chandrasekhar}{1943}]{cha43}
  Chandrasekhar S., 1943, ApJ, 97, 255

\bibitem[\protect\citeauthoryear{De Propris et al.}{2005}]{dep05} 
  De Propris R., Phillipps S., Drinkwater M.J., Gregg M.D., Jones
  J.B., Evstigneeva E., Bekki K., 2005, ApJ, 623, 105L

\bibitem[\protect\citeauthoryear{Di Matteo et al.}{2003}]{dim03} 
  Di Matteo T., Allen S.W., Fabian A.C., Wilson A.S., Young A.J.,
  2003, ApJ, 582, 133

\bibitem[\protect\citeauthoryear{Dirsch et al.}{2003}]{dir03} 
  Dirsch B., Richtler T., Geisler D., Forte J.C., Bassino L.P., Gieren
  W.P., 2003, AJ, 125, 1908

\bibitem[\protect\citeauthoryear{Evstigneeva, Gregg \& Drinkwater}
  {Evstigneeva et al.}{2005}]{evs05} 
  Evstigneeva E.A., Gregg M.D., Drinkwater M.J., 2005, to appear in
  the proccedings of IAU Colloquium 198, astro-ph/0504289

\bibitem[\protect\citeauthoryear{Fellhauer \& Kroupa}{2002}]{fel02}
  Fellhauer M., Kroupa P., 2002, MNRAS, 330, 642

\bibitem[\protect\citeauthoryear{Fellhauer \& Kroupa}{2005}]{fel05}
  Fellhauer M., Kroupa P., 2005, MNRAS, 359, 223

\bibitem[\protect\citeauthoryear{Fellhauer et al.}{2000}]{fel00}
  Fellhauer M., Kroupa P., Baumgardt H., Bien R., Boily C.M., Spurzem
  R., Wassmer N., 2000, NewA, 5, 305

\bibitem[\protect\citeauthoryear{Hasegan et al.}{2005}]{has05} 
  Hasegan M., Jordan A., Cote P., Djorgovski S.G., Mclaughlin D.E.,
  Blakeslee J.P., Mei S., West M.J., Peng E.W., Ferrarese L.,
  Milosavljevic M., Tonry J.L. Merrit D., 2005, ApJ, 627, 203

\bibitem[\protect\citeauthoryear{Hilker}{1998}]{hil98}
  Hilker M., 1998, PhD-thesis, University of Bonn

\bibitem[\protect\citeauthoryear{Hilker, Infante \& Richtler}{Hilker
    et al.}{1999}]{hil99} 
  Hilker M., Infante L., Richtler T., 1999, A\&AS, 138, 55

\bibitem[\protect\citeauthoryear{King}{1966}]{kin66} 
  King I., 1966, AJ, 71, 61

\bibitem[\protect\citeauthoryear{Kroupa}{1997}]{kro97} 
  Kroupa P., 1997, NewA, 2, 139

\bibitem[\protect\citeauthoryear{Kroupa}{1998}]{kro98} 
  Kroupa P., 1998, MNRAS, 300, 200

\bibitem[\protect\citeauthoryear{Maraston et al.}{2004}]{mar04}
  Maraston C., Bastian N., Saglia R.P., Kissler-Patig M., Schweizer
  F., Goudfrooij P., 2004, A\&A, 416, 467

\bibitem[\protect\citeauthoryear{Mayer et al.}{2001}]{may01}
  Mayer L., Governato F., Colpi M., Moore B., Quinn T., Wadsley J.,
  Stadel J., Lake G., 2001, ApJ, 559, 754

\bibitem[\protect\citeauthoryear{Mayer et al.}{2002}]{may02}
  Mayer L., Moore B., Quinn T., Governato F., Stadel J., 2002, MNRAS,
  336, 119

\bibitem[\protect\citeauthoryear{McLaughlin}{1999}]{mcl99} 
  McLaughlin D.E., 1999, ApJ, 512, 9L

\bibitem[\protect\citeauthoryear{Mieske, Hilker \& Infante}{Mieske et
    al.}{2002}]{mie02} 
  Mieske S., Hilker M., Infante L, 2002, A\&A, 383, 823

\bibitem[\protect\citeauthoryear{Mieske et al.}{2004}]{mie04} 
  Mieske S., Infante L., Benítez N., Coe D., Blakeslee J.P., Zekser
  K., Ford H.C., Broadhurst T.J., Illingworth G.D., Hartig G.F.,
  Clampin M., Ardila D.R., Bartko F., Bouwens R.J., Brown R.A.,
  Burrows C.J., Cheng E.S., Cross N.J.G., Feldman P.D., Franx M.,
  Golimowski D.A., Goto T., Gronwall C., Holden B., Homeier N.,
  Kimble, R.A., Krist J.E., Lesser M.P., Martel A.R., Menanteau F.,
  Meurer G.R., Miley G.K., Postman M., Rosati P., Sirianni M., Sparks
  W.B., Tran H.D., Tsvetanov Z.I., White R.L., Zheng W., 2004, AJ,
  128, 1529 

\bibitem[\protect\citeauthoryear{Phillipps et al.}{2001}]{phi01}
  Phillipps S., Drinkwater M.J., Gregg M.D., Jones J.B., 2001, ApJ,
  560, 201

\bibitem[\protect\citeauthoryear{Plummer}{1911}]{plu11} 
  Plummer H.C., 1911, MNRAS, 71, 460

\bibitem[\protect\citeauthoryear{Portegies Zwart \& McMillan}{2002}]
  {por02} 
  Portegies Zwart S.F., McMillan S.L.W., 2002, ApJ, 576,899

\bibitem[\protect\citeauthoryear{Vesperini et al.}{2003}]{ves03}
  Vesperini E., Zepf S.E., Kundu A., Ashman K.M., 2003, ApJ, 593, 760

\end{thebibliography}
\end{document}